\documentclass[a4paper]{amsart}

\usepackage{amsmath}  
\usepackage{amsfonts}
\usepackage{amssymb}
\usepackage{times}
\usepackage{helvet}
\usepackage{bbm}

\usepackage{afterpage}

\usepackage{array}
\usepackage{graphicx}
\usepackage{color}

\usepackage[swedish,english]{babel}

\usepackage[latin1]{inputenc}
\usepackage[T1]{fontenc}

\usepackage{float}
\usepackage{tikz}
\usetikzlibrary{arrows,decorations.pathmorphing,matrix,chains,positioning}

\theoremstyle{plain}
\newtheorem{thm}{THEOREM}[section]
\newtheorem{lm}[thm]{LEMMA}

\newtheorem{prop}[thm]{PROPOSITION}

\theoremstyle{definition}

\newtheorem{remark}[thm]{Remark}

\theoremstyle{remark}

\newcommand{\R}{\mathbb{R}}

\newcommand{\Z}{\mathbb{Z}}

\newcommand{\EE}{\mathbb{E}}

\newcommand{\Pe}{\mathbb{P}}
\newcommand{\bigoh}{\mathcal{O}}

\newcommand{\one}{\mathbbm{1}}
\newcommand\tthash{\text{\ttfamily\#}}

\newcommand{\calx}{{\mathcal X}}
\newcommand{\calxe}{{\mathcal X}_{\epsilon}}
\newcommand{\calye}{{\mathcal Y}_{\epsilon}}
\newcommand{\calyee}{{\mathcal Y}_{\epsilon,\epsilon}}

\newcommand{\tT}{{\widetilde{T}}}
\newcommand{\tV}{{\widetilde{V}}}
\newcommand{\taueps}{{\tau_{\epsilon}}}
\newcommand{\ttau}{\tilde{\tau}}
\newcommand{\tf}{\tilde{f}}

\newcommand{\bart}{\bar{t}}

\title{The Lorentz gas with a nearly periodic distribution of scatterers}
\author{Bernt Wennberg$^{(1)}$ }

\newcommand*{\allpages}{%
    \let\emptyevenpage\relax
}

\begin{document}

\maketitle

\begin{center}
(1) Department of Mathematical Sciences, \\
Chalmers University of Technology and University of Gothenburg\\
SE 41296 G\"oteborg, Sweden \\
email: wennberg@chalmers.se

\end{center}

\begin{abstract} We consider the Lorentz gas in a distribution of
  scatterers which microscopically converges to a periodic
  distribution, and prove that the Lorentz gas in the low density
  limit satisfies a linear Boltzmann equation. This is in contrast
  with the periodic Lorentz gas, which does not satisfy the Boltzmann
  equation in the limit. 
\end{abstract} 


\section{Introduction}
\label{sec:intro}

The Lorentz gas was introduced in~\cite{Lorentz1905} to give a new
understanding of phenomena such as electric resistivity and the Hall
effect. Lorentz introduced many simplifications to admit ``rigorously
exact solutions'' to some questions, which has made the model very
attractive in the mathematical community. 

\begin{figure}[h]
  \centering
  \includegraphics[width=0.6\textwidth]{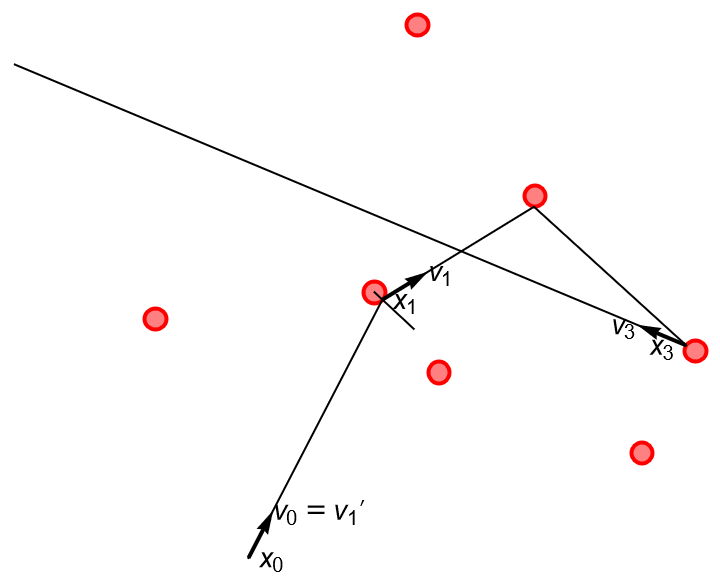}
  \caption{A trajectory of the Lorentz process}
  \label{fig:fig1}
\end{figure}

The Lorentz gas can be described as follows: Let $\calx$ be a point set
in $\R^2$ (actually most of what is said in this paper could equally
well have been set in $\R^n$ with $n\ge 2$). The point set can be
expressed as a locally finite counting measure,
\begin{equation}
  \label{eq:001}
  \calx(A) = \tthash\{ X_i\in A\}\,,
\end{equation}
the number of points in the set $A$, or the empirical measure
\begin{equation}
  \label{eq:005}
  \calx = \sum \delta_{X_j}\,,
\end{equation}
Examples of interest  are the periodic set $\calx=\Z^2$, or
random point processes such as a Poisson distribution. The Lorentz
process is the motion of a point particle with constant speed in the
plane with an obstacle of a fixed radius $r$ at each point
$X\in\calx$. Given an initial position and velocity of the point
particle, $(x_0,v_0)\in \R^2\times S^1$, its position at time $t$ is 
given by
\begin{align}
  \label{eq:006}
  (x(t),v(t)) &= T^t_{\calx,r}( (x_0,v_0)) = \Big(x_0 + \sum_{j=1}^M
                (t_j-t_{j-1}) v_{j-1} + (t-t_M) v_M, v_M\Big)\,,
\end{align}
where $t_0=0$, and $\{t_1,...,t_M\}$ is the set of times where the trajectory of
the point particle hits an obstacle, and where $v_j$ is the new  
velocity that results from a specular reflection on the obstacle. We
also set $x_j=x(t_j)$. The 
notation is clarified in Fig.~\ref{fig:fig1}.  For
a fixed point set $\calx$ this is a deterministic motion, but when
$\calx$ is random, so is the motion of the point particle.

For a given point set $\chi$ we also consider the rescaled set
$\calxe=\sqrt{\epsilon} \calx$, so that for any set $A\subset\R^2$ 
\begin{align}
  \calxe(A) = \calx(\frac{A}{\sqrt{\epsilon}})\,.
\end{align}
We think of $\chi$ as describing the domain of the Lorentz gas at a
microscopic scale, and $\chi_{\epsilon}$ as the macroscopic
scale. Expressed in the macroscopic scale, we will assume that for any
open set $A\subset \R^2$
\begin{equation}
  \label{eq:008}
  \lim_{\epsilon\rightarrow 0} \epsilon\;  \tthash\{ x\in{\chi}_{\epsilon}\cap A\} = c m(A)\,,
\end{equation}
where $m(A)$ is the Lebesgue measure of
the set $A$ and $c$ is a positive constant. Some additional condition
is needed to  ensure that the motion is (almost 
certainly) well defined, and that the number of encounters with
obstacles, $M$, is finite. This is true, for example, if the point set
is a Delone set, and hence satisfies bounds on the minimal distance
between the points as well as on the density of points.  

For the rest of the paper
the obstacle radius is fixed to be equal to $\epsilon$ in the
macroscopic scaling, and therefore $\sqrt{\epsilon}$ in the
microscopic scaling.

 In the macroscopic scale, the time $t_1$ of the first encounter with
an obstacle for  a typical trajectory $T_{\calxe,
  \epsilon}^t(x_0,v_0)$ satisfies $t_1 =\bigoh(1/c)$, {\em i.e.} the
{\em mean free path-length} of a typical trajectory is of the order
$1/c$. This is known as the low density limit, or the Boltzmann-Grad
limit.

Consider next an initial  density of point particles
$f_0\in L^1_+(\R^2\times S^1)$, and its evolution under the Lorentz
process. The density at a later time $t$ is then $f_{\epsilon,t}=f_{\epsilon}(x,v,t) = \EE[
f_0(T^{-t}_{\calxe,\epsilon}(x,v))]$, where the expectation is taken
with respect to the probability distribution of the point set $\calxe$
in case it is random.

Equivalently (because for a fixed realization of
$\calxe$, the orbit $T^t_{\calxe,\epsilon}$ is reversible), for each
$g\in C_0(\R^2\times S^1)$ 
\begin{equation}
  \label{eq:020}
  \int_{\R^2\times S^1} f_{\epsilon}(x,v,t) g(x,v) \,dx dv =  \int_{\R^2\times
    S^1} f_0(x,v) \EE[ g( T^t_{\calxe,\epsilon}(x,v)) ] \,dx dv.
\end{equation}

Gallavotti~\cite{Gallavotti1972,Gallavotti1999book} proved that when
$\calx$ is a Poisson process with unit 
intensity, and $\epsilon$ converges to zero, then $f_{\epsilon,t}$
converges to a density $f_t$ which satisfies the linear Boltzmann
equation: 
\begin{equation}
  \label{eq:025}
  \partial_t f(x,v,t) + v\cdot \nabla_x f(x,v,t)  = -2 f(x,v,t)
  +  \int_{S^1_-} f(x,v',t) |v\cdot \omega|\,d\omega\,.
\end{equation}
Here $S^1_- =S^1_-(v) =  \{\omega\in S^1\;| \; v\cdot\omega <0 \}$ and
$v'= v-2(\omega,v)\omega$. Spohn has proven a related, and more
general,  result in 
\cite{Spohn1978}. 

On the other hand, it is also known that
when $\calx=\Z^2$ (or for that matter many other regular point sets,
such as quasi crystals), then the Lorentz process is not Markovian in
the limit and therefore~(\ref{eq:020}) does not hold, see
\cite{BourgainGolseWennberg1998,Golse2008}. For a periodic
distribution of scatterers, Caglioti and Golse
\cite{CagliotiGolse2010} proved that there is a
limiting kinetic equation in an enlarged phase space, and more general
results of the same kind were obtained by Marklof and
Strömbergsson~\cite{  MarklofStrombergsson2008, MarklofStrombergsson2010}. 

Marklof and Strömbergsson has studied this problem in several papers
\cite{MarklofStrombergsson2011,MarklofStrombergsson2011b,
  MarklofStrombergsson2014,MarklofStrombergsson2015},  
and present in \cite{MarklofStrombergsson_arxiv2019} a very general theorem
concering the Boltzmann-Grad limit of the Lorentz process, where  they
give a  concise set of conditions for a point  set $\calx$, such
that the Lorentz process $T^t_{\calxe,\epsilon}(x,v)$ converges to a
free flight process. The point set $\calx$ here is a fixed point set,
that could be for example a periodic set, or  one given realization of
a random point process, and hence, for a fixed $\epsilon$ the particle
flow is deterministic and not the expectation over the point process
as in Eq.~(\ref{eq:020}). 

The problem  studied in this paper is the following: Let $\calx$ be a
given point process, and let $\calye$ be a family of point processes
converging to $\calx$ in the sense that the Lévy-Prohorov distance
$\pi(\calx,\calye)$ between the measures~$\calye$ and $\calx$
converges to zero ( a.s. ) when $\epsilon\rightarrow 0$. Recall here
the definition of the Lévy-Prohorov distance:
\begin{equation}
  \label{eq:030}
  \pi(\nu,\mu) = \inf \left\{ \varepsilon >0 \; \bigg|
    \mu(A) \le \nu(A^{\varepsilon})+\varepsilon  \;\mbox{and}\;
    \nu(A) \le \mu(A^{\varepsilon})+\varepsilon  \; \mbox{for all} \; A\in\mathcal{B}\right\}\,.
\end{equation}
Here $\mathcal{B}$ is the Borel sigma algebra (of $\R^2$ in our case),
and
\begin{equation}
  \label{eq:030b}
  A^{\varepsilon} = \bigcup_{x\in A} B_{\varepsilon}(x)
\end{equation}
where $B_{\varepsilon}(x)$ is the ball of radius $\varepsilon$ and
centre at $x\in A$.  This convergence is thus assumed to
take place at the microscopic level. To study this in the
Boltzmann-Grad limit, we set 
\begin{equation}
  \label{eq:031}
  \calyee = \sqrt{\epsilon}\calye\qquad\mbox{and}\qquad\calxe
  =\sqrt{\epsilon} \calx\,.
\end{equation}
We are then interested in  comparing the limits of the corresponding
Lorentz processes, $T_{\calyee}^t$ and  
$T_{\calxe}^t$, assuming that the obstacle radius is $\epsilon$. The
main result of the paper is the construction of a family of point sets
$\calye$ that converges to the periodic distribution, and yet the
Lorentz process 
$T_{\calyee}^t(x,v)$ converges to the free flight process generated by
the linear Boltzmann equation~(\ref{eq:020}), contrary to the limit of
$T_{\calxe}^t(x,v)$. It is in a sense a non-stability result for the
periodic Lorentz gas, and while not proven in this paper it seems very
likely that if $\calx$ is a Poisson process with constant intensity,
(almost) any approximation $\calye$ would result in convergence of the
two Lorentz processes to the same limit. 
The proof follows quite closely the construction
in~\cite{CagliotiPulvirentiRicci2000} and~\cite{RicciWennberg2004},
and consists in constructing a third process, $\tT_{\calyee}^t$
which which can be proven to be path-wise close the free flight
process, and also to the Lorentz process.  The full statement of the
result, together with the main steps of the proof are given in
Section~\ref{sec:main}. Section~\ref{sec:markovian} gives the somewhat
technical proof that $\tT_{\calyee}^t$ converges to the
Boltzmann process, and Section~\ref{sec:equivalence} contains a proof
that the probability that an orbit of $T_{\calyee}^t$ crosses itself
near an obstacle is negligible in the limit as $\epsilon\rightarrow 0$,
which then used to prove that $\tT_{\calyee}^t$ and
$T_{\calyee}^t$ with large probability are path-wise close.

The scaling studied here for the Lorentz process is not the only one
studied in literature. A more challenging problem is the long time
limit, where the process is studied over a time interval of
$[0,t_\epsilon[$, where $t_\epsilon\rightarrow\infty$ when
$\epsilon\rightarrow 0$. Recent results of this kind have been obtained
in~\cite{LutskoTot2020}. Other related results can be found
in~\cite{PeneTerhesiu2020} and~\cite{MelbournePeneTerhesiu2021}.

\section{The main result and the principal steps of its proof}
\label{sec:main}

Let $\calx=\Z^2$ and define $\calye$ as a perturbation of $\calx$ in
the following way: Let $\phi$ be a rotationally symmetric probability
density  supported in 
$|x| < 1$, fix $\nu\in]1/2,1[$, set
\begin{equation}
  \label{eq:2010}
  \calye = \left\{ (j,k) +\epsilon^{1-\nu} \xi_{j,k}\;|\; (j,k)\in\Z^2,
    \xi_{j,k} \;\mbox{i.i.d. with density}\; \phi\right\}\,.
\end{equation}
and let
\begin{equation}
  \label{eq:2011}
  \calyee = \sqrt{\epsilon}\calye\,.
\end{equation}
Thus each obstacle has its center in a disk of radius
$\epsilon^{1-\nu}$ centered at an integer coordinate, $(j,k)$. This
disk will be called the {\em obstacle patch} below. We have almost
trivially that the Lévy-Prohorov distance between $\calx$ and $\calye$
is smaller than $\epsilon^{1-\nu}$ for all realizations of the random
set $\calye$. The obstacle itself reaches at most a distance
$\epsilon^{1-\nu} + 
\epsilon^{1/2}$ from the same integer coordinate; this larger disk
will be called an {\em obstacle range} below. The ratio of the
obstacle range and the support of the center distribution is thus
$1+\epsilon^{\nu-1/2}$, which converges to $1$ when
$\epsilon\rightarrow 0$, and to simplify some notation the radius of
the obstacle patch will be used instead of the radius of the obstacle
range, and the difference will be accounted for with a constant in the
estimates. 

Clearly $\calye$ converges in law to the periodic distribution in
$\R^2$. Nevertheless we have the following theorem:

\begin{thm}
  \label{thm:main}
  Let $T_{\calyee}^t(x,v)$ be the Lorentz process obtained by placing
  a circular obstacle of radius $\epsilon$ at each point of
  $\calyee$. Let $\bart>0$, and let  $f_0(x,v)$ be a  probability density in
  $\R^2\times S^1$.  Define  $f_{\epsilon}(x,v,t)$ as  the function such
  that for all $t\in [0, \bart]$
  $g\in C(\R^2\times S^1)$,

  \begin{equation}
    \label{eq:2020}
    \int_{\R^2\times S^1} f_{\epsilon}(x,v,t) g(x,v)\,dx dv =
    \int_{\R^2\times S^1} f_0(x,v) \EE[ g(T_{\calyee}^t(x,v)) ]\,dxdv\,.
  \end{equation}

  Then there is a density $f(x,v,t)\in C([0,\bart],L^1(\R^2\times
  S^1)$ such that for all $t\le \bart$, $f_\epsilon(x,v,t)\rightarrow f(x,v,t)$ in
  $L^1(\R^2\times S^1)$ when $\epsilon\rightarrow 0$, and such that  $f(x,v,t)$
  satisfies the linear Boltzmann equation
  \begin{equation}
    \label{eq:2021}
    \partial_t f(x,v,t) + v\cdot \nabla_x f(x,v,t) = \kappa \left(
       \int_{S^1_-} f(x,v',t) |v\cdot\omega| \,d\omega - 2 f(x,v,t)
    \right)\,.
  \end{equation}
  The constant $\kappa$ depends on $\phi$ and $S^1_-$ and $v'$ are
  defined as in Eq.~(\ref{eq:025}).
\end{thm}

\begin{remark}
  To simplify notation, point particles are allowed to start inside
  obstacles, and to cross the obstacle boundary from the
  inside without any change of velocity.
\end{remark}

Both the statement in this theorem and the proof are very similar to
the main results of~\cite{CagliotiPulvirentiRicci2000} and
\cite{RicciWennberg2004}, but the implication is anyway quite
different. In those papers the point processes $\calye$ are
constructed as the thinning of a periodic point set:
\begin{equation}
  \label{eq:2030}
  \calye = \left\{ \epsilon^{\nu}(j,k)\; |\; \xi_{j,k}=1, \; \xi_{j,k}
    \;\mbox{i.i.d. Bernoulli with } \Pe[\xi_{j,k}=1]=\epsilon^{2\nu}\right\}\,.
\end{equation}
Here it is clear that this $\calye$ converges in law to the Poisson process
with intensity one, and in hindsight it is perhaps not surprising that
the limit of the Lorentz process is the same as for the Lorentz
process generated by a Poisson distribution of the obstacles.
Theorem~\ref{thm:main} in the present paper states that
the limit of the Lorentz process is the free flight process of the
Boltzmann equation in some cases also  if the limiting obstacle
density is periodic, and raises the question as to which point
processes $\calx$ are stable to perturbation when it comes to the low
density limit of the corresponding Lorentz processes.

In the proof we consider three processes: The Lorentz process
$T_{\calyee}^t(x,v)$, the free flight process $T_B^t(x,v)$ generated
by the Boltzmann equation~(\ref{eq:2020}), and an auxiliary Markovian
Lorentz process, $\tT_{\calyee}^t(x,v)$.

{\em The Boltzmann process} is the random flight process
$(x(t),v(t))=T^t_B(x,v) $  generated by 
Eq.~(\ref{eq:001}). Let $0=t_0<t_1<,...,t_n<t$ be a sequence of times
generated by independent, exponentially distributed increments
$t_j-t_{j-1}$ with intensity $2$, let $v_0=v$ define $v_j=v_{j-1}-2
(\omega_j\cdot v_{j-1})\,\omega_j$, where the $\omega_j\in S^1$ are
independent and uniformly distributed. Finally set
\begin{align}
  \label{eq:2035}
  x(t)& = x_n(t) = x+t_1 v+ (t_2-t_1)v_1+\cdots + (t_n-t_{n-1})v_{n-1} +
        (t-t_n)v_n \nonumber \\
  v(t)&= v_n\,.
\end{align}
Of course the number $n$ in the sum is then Poisson distributed. The
solution of Eq.~(\ref{eq:001}) may be defined weakly by
\begin{equation}
  \label{eq:2036}
  \int_{\R^2\times S^1} f_0(x,v) g(x,v,t)\,dx dv = \int_{\R^2\times
    S^1}f(x,v,t) g(x,v)\,dx dv\,.
\end{equation}
The function $g(x,v,t)$ can then  be expanded as a sum of terms, each representing
the paths with a fixed number of jumps:
\begin{equation}
  \label{eq:2037}
  g(x,v,t) = V^t g(x,v) = \sum_{n=0}^{\infty} g_n(x,v,t)\,,
\end{equation}
with
\begin{align}
  \label{eq:2038}
  g_n(x,v,t)  = (V^tg)_n(x,v)= & e^{-2 t}
  \int_{0}^{t}\int_{t_1}^t\cdots\int_{t_{n-1}}^t  dt_1\cdots dt_n \times \nonumber \\
  & \int_{(S^1_-)^n}d\omega_1\cdots d\omega_n \prod_{k=1}^n |
  \omega_k\cdot v_{k-1}| g(x_n(t), v_n)\,.
\end{align}

\begin{figure}[h]
  \centering
  \includegraphics[width=0.6\textwidth]{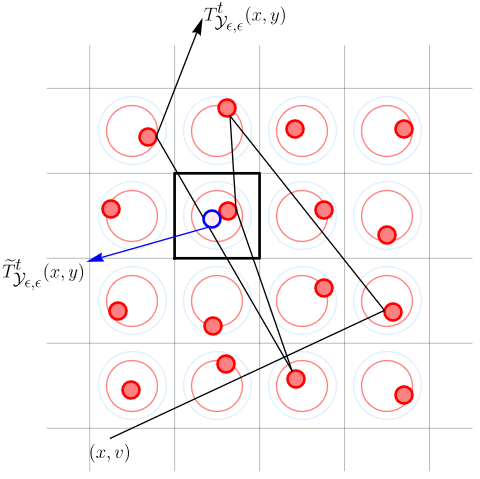}
  \caption{Trajectories for the Lorentz process and the Markovian
    Lorentz process. The obstacle size and the support for the
    probability distributions are exaggerated.}
  \label{fig:fig2}
\end{figure}

{\em The Markovian Lorentz process} here is  similar to the
corresponding process
in~\cite{CagliotiPulvirentiRicci2000,RicciWennberg2004}, in the way
that  re-collisions, {\em i.e.} the event that a particle
trajectory meets the same obstacle a second time, are eliminated. The
construction is explained in Fig.~\ref{fig:fig2}. The red thin circles
mark the support of the distribution of obstacle center in each
cell, and the blue circle indicates the reach of an obstacle, the
obstacle range.  In each cell there is an obstacle, fixed from the
start in the Lorentz model, but changing in the Markovian Lorentz
model. Note that the size of both the obstacle radius and the obstacle
support are very small, and decreasing to zero with $\epsilon$, but
drawn large here, for clarity. The path meets the same obstacle
patch twice in the boxed cell. In the Lorentz case, the
orbit simply traverses the cell the second time, because it misses the
obstacle. In the Markovian case the obstacle is at a new, random
position,  drawn in blue color, the second time the orbit enters the 
support, and there is a  positive probability that the path collides
with the obstacle, as shown with the blue trajectory. In both cases
the trajectory is surely well defined, but the probability of
realizing an orbit is different.

The process can be defined as in Eq.~(\ref{eq:2036}), with the function
$g(x,v,t)$ replaced by
\begin{align}
  \tilde{g}_{\epsilon}(x,v,t) 
   &=
  \EE\left[g(\tT_{\calyee}^t(x,v))\right]=
  \sum_{n=0}^{\infty}\tilde{g}_{\epsilon,n}(x,v,t)\,,\\
  \left(\tV^t_{\epsilon,n}\right) g(x,v) &= \tilde{g}_{\epsilon,n}(x,v,t)
\end{align}
where $\tilde{g}_{\epsilon,n}(x,v,t)$ is the contribution of
trajectories having exactly $n$ velocity jumps in the interval  $[0,t]$. All
these terms can be computed rather explicitly by counting the number
of times a trajectory crosses the obstacle range of one
cell.

To simplify notation these three
processes are hereafter denoted $z_{\epsilon}(t)$, $z(t)$ and
$\tilde{z}_{\epsilon}(t)$. As described
in~\cite{CagliotiPulvirentiRicci2000,RicciWennberg2004} all these
processes belong with probability one to the Skorokhod space
$D_{[0,\bart]}(\R^2\times S^1)$ of {\em càdlàg} functions on
$\R^2\times S^1$ equiped with the distance
\begin{equation}
  \label{eq:2040}
  d_S(x,y) = \inf_{\lambda\in\Lambda}\left\{
  \sup_{t\in[0,\bart]}\| x(t)-y(\lambda(t))\|_{\R^2\times S^1}+
    \sup_{t\in[0,\bart]}|t-\lambda(t)|  \right\}\,,
\end{equation}
where
\begin{equation}
  \label{eq:2041}
  \Lambda=\{\lambda\in C([0,\bart]): t>s \implies
  \lambda(t)>\lambda(s), \lambda(0)=0, \lambda(\bart)=\bart \}\,.
\end{equation}

Any $z\in C(D_{[0,\bart]}(\R^2\times S^1))$ induces a measure $\mu$
on $D_{[0,\bart]}(\R^2\times S^1)$
which is first defined on cylindrical continuous functions $F$, {\em i.e.} 
 functions  $F\in C(D_{[0,\bart]}(\R^2\times S^1))$ of the form
 $F(z) = F_n(z(t_1),z(t_2),...,z(t_n))$ where 
$F_n\in C( (\R^2\times S^1)^n)$ and $0\le t_1<t_2<\cdots <t_n\le
\bart$. For such functions  $\mu$ is defined by
\begin{equation}
  \label{eq:2042}
  \int F(z) \mu(dz) = \int f_0(z)
  P_{t_n,...,t_1,0}(z_1,z_2,\cdots,z_n| z_0)dz_0 dz_1 \cdots dz_n\,,
\end{equation}
where $P_{t_n,...,t_1,0}(z_1,z_2,\cdots,z_n| z_0)$ is the joint probability
density of $z(t_1),z(t_2),...,z(t_n)$ given the starting point
$z_0$. For a Markov process, such as the Boltzmann process, this is
\begin{equation}
  \label{eq:2042b}
  \int F(z) \mu(dz) = \int f_0(z)
  P_{t_1,0}(z_1|z_0)P_{t_2,t_1}(z_2|z_1)\cdots
  P_{t_n,t_{n-1}}(z_n|z_{n-1})dz_0 dz_1 \cdots dz_n\,,
\end{equation}
where $P_{t_j,t_{j-1}}(z_j | z_{j-1})$ is the transition probability
of going from state $z_{j-1}$ to state $z_j$ in the time interval from
$t_{j-1}$ to $t_j$. Then by a density argument the measure $\mu$ is defined for
all $F\in C(D_{[0,\bart]}(\R^2\times S^1))$.

\medskip
\noindent{\bf Proof:} (of Theorem~\ref{thm:main})
Let $\mu$, $\mu_{\epsilon}$ and $\tilde{\mu}_{\epsilon}$ be the
measures induced by $z(t)$, $z_{\epsilon}(t)$ and
$\tilde{z}_{\epsilon}(t)$. Just like in
in~\cite{RicciWennberg2004}, one may prove that for every $F$ 
\begin{equation}
  \label{eq:2050}
  \lim_{\epsilon\rightarrow 0} \int F(z) \tilde{\mu}_{\epsilon}(dz)
  \rightarrow  \int F(z) \mu(dz)\,.
\end{equation}
The argument uses a theorem from Gikhman and Shorokhod%
(1974)
\cite{GikhmanSkorokhod1974}, and relies on an
equicontinuity condition and on the convergence of the marginal
distributions of $\tilde{z}_{\epsilon}(t)$. Both the equicontinuity condition
and the convergence of the marginal distributions are consequences of
Proposition~\ref{prop:V2} which shows that  the number of jumps of
$z_{\epsilon}$ in small time intervals is not too large, and Theorem~\ref{thm:MarkovToBoltzmann},  which states that the one-dimensional marginals of
$\tilde{z}_{\epsilon}(t)$ converge to the marginals of $z(t)$.

The next step is to prove that
$(\mu_{\varepsilon}-\tilde{\mu}_{\epsilon} )\rightharpoonup 0$ when
$\epsilon\rightarrow 0$, {\em i.e.}
\begin{equation}
  \label{eq:2060}
  \left| \int F(z) \mu_{\epsilon}(dz)
  - \int F(z) \tilde{\mu}_{\epsilon}(dz)\right|\rightarrow 0
\end{equation}
when $\epsilon\rightarrow 0$, and this is done by a coupling
argument. Given that $z_{\epsilon}(0)=\tilde{z}_{\epsilon}(0)$,
the two processes have  the same marginal distributions up to
the first time $t_*$ where
$z_{\epsilon}(t)=\tilde{z}_{\epsilon}(t)$ returns to an obstacle range
it already visited. For the process $z_{\epsilon}(t)$, the position 
of the obstacle within its support is then fixed, whereas for the
process $\tilde{z}_{\epsilon}(t)$ a new random position of the obstacle
is chosen when the trajectory arrives. Hence 
\begin{align}
  \label{eq:2061}
  \left| \int F(z) \mu_{\epsilon}(dz)
  - \int F(z) \tilde{\mu}_{\epsilon}(dz)\right| &=  \left| \int_{K_{\epsilon}}
                                       F(z) \mu_{\epsilon}(dz) 
  - \int_{K_{\epsilon}} F(z) \tilde{\mu}_{\epsilon}(dz)\right|
                                       \nonumber \\
 &\le \sup |F|
\left(\mu_{\epsilon}(K_{\epsilon}) + \tilde{\mu}_{\epsilon}(K_{\epsilon})\right)\,,
\end{align}
where $K_{\epsilon}\subset D_{[0,\bart]}(\R^2\times S^1)$ is the set
of trajectories that contain at least one such re-encounter for $t\le
\bart$. Proposition~\ref{prop:noloops}  states that
the righthand side of (\ref{eq:2061}) converges to zero when $\epsilon
\rightarrow 0$. This together with (\ref{eq:2050}) implies that
\begin{equation}
  \label{eq:2070}
  \mu_{\epsilon}\rightarrow \mu
\end{equation}
weakly when $\epsilon\rightarrow 0$, and this concludes the proof of
Theorem~\ref{thm:main}.  \hfill $\square$

\medskip

We end the section with a comment on the propagation of chaos for
these procesess. For the Boltzmann process $z(t)$ and for the process
$\tilde{z}(t)$ it is clear that if a pair of initial conditions
$(z^1(0),z^2(0))$ are chosen with joint density $f_0^1(x,v)
f_0^2(x,v)$, then joint densities of $(z^1(t),z^2(t))$ and
$(\tilde{z}_{\epsilon}^1(t),\tilde{z}_{\epsilon}^2(t))$ also
  factorize: a chaotic initial state is propagated by the flow,
  because the two paths are independent, they never interact. The same
  is not true for
  $(z_{\epsilon}^1(t),z_{\epsilon}^2(t))$, because
  there is a positive probability that the two paths will meet in the
  obstacle density support of one obstacle, which creates
  correlations. However, the same kind of estimates as the ones used
  to prove that the probability of re-encounters for one trajectory
  becomes negligible when $\epsilon\rightarrow 0$ can be used to prove
  that  also the probability that two different trajectories meet
  inside an obstacle range
  becomes small, and such estimates could be performed for any finite
  number of trajectories. The calculations are carried out in some
  detail in Section~\ref{sec:equivalence}, and formlated as
  Theorem~\ref{thm:chaos}.   

\section{The Markovian Lorentz process}
\label{sec:markovian}

The Markovian process may be described using the underlying periodic
structure. Over a time interval $[0,t]$, the particle traverses
$\bigoh\left(\epsilon^{-1/2}\right)$ lattice cells, and when the path
is sufficiently close to the cell center to cross the obstacle range,
there is a positive 
probability that it is reflected by an obstacle. Because the
obstacle position is chosen independently each time the trajectory
enters a cell, this may all be computed rather explicitly.

We write
\begin{equation}
  \label{eq:3020}
  \tilde{g}_{\epsilon}(x,v,t) = 
  \EE\left[g(\tT_{\calyee}^t(x,v))\right]= \widetilde{V}_{\epsilon}^tg(x,v)
 =  \sum_{n=0}^{\infty}  (\widetilde{V}_{\epsilon})_n^tg(x,v) \,,
\end{equation}
This is the test function evaluated along the path of a point
particle. It defines a semigroup $\widetilde{V}_{\epsilon}^t$ acting on
the test function, and the terms in the sum express the contribution to
this semigroup from paths with exactly $n$ velocity jumps. Obviously
these terms are not semigroups in their own right.

\begin{thm}
  \label{thm:MarkovToBoltzmann}
Let $\tT_{\calyee}^t(x,v)$ be the Markovian Lorentz process as defined
above. Fix $\bart>0$. For any density $f_0(x,v)$ in $\R^2\times
S^1$, let $\tf_{\epsilon}(x,v,t)$ be the unique function such that 
\begin{equation}
  \label{eq:thm3_1}
  \int_{R^2\times S^1} \tf_{\epsilon}(x,v,t) g(x,v)\,dxdv =
  \int_{R^2\times S^1} f_0(x,v) g(\tT_{\calyee}^t(x,v))\,dxdv\,,
\end{equation}
for all  $t\in[0,\bart]$, and any $g(x,v)\in C(\R^2\times
S^1)$. Then there is a function $f(x,v,t)$ such that for all $t$
\begin{equation}
  \label{eq:thm3_2}
  \tf_{\epsilon}(x,v,t)\rightarrow f(x,v,t)\,,
\end{equation}
and $f(x,v,t)$ satisfies the linear Boltzmann equation , Eq.~(\ref{eq:025}).
\end{thm}

\noindent{\bf Proof:}
Let $g\in C_0(\R^2\times S^1)$, and $t<\bart$. We must prove that
\begin{equation}
  \label{eq:3050}
  \int_{\R^2\times S^1} \left( \tilde{f}_{\epsilon}(x,v,t)-f(x,v,t)\right)
  g(x,v)\,dxdv\rightarrow 0\quad \mbox{when} \quad
  \epsilon\rightarrow 0\,,
\end{equation}
or, equivalently,
\begin{equation}
  \label{eq:3051}
  \int_{\R^2\times S^1} f_0(x,v)
  \left(V^tg(x,v)-\tilde{V}_{\epsilon}^tg(x,v)\right) \,dxdv
  \rightarrow 0\,.
\end{equation}
The expression in (\ref{eq:3051}) is bounded by
\begin{equation}
  \label{eq:3055}
  \|g\|_{L^{\infty}} \int_{\R^2\times S^1}
  f_0\left(\one_{f_0>\lambda} + \one_{|x|\ge M}\right)  \, dxdv + \lambda \|
  (V^tg-\tilde{V}_{\epsilon}^t g) \one_{|x|< M}\|_{L^1}\,,
\end{equation}
and choosing $\lambda$ and $M$ large enough, depending on $f_0$,  the
first term can 
be made arbitrarily 
small, smaller than $\varepsilon_0/2$, say, where $\varepsilon_0$ is
taken arbitrarily small. It the remains to show that the second term
can be made smaller than $\varepsilon_0/2$ by choosing $\epsilon$ small
enough.  
Both $V^t$ and $\tV_{\epsilon}^t$ are bounded semigroups, and
therefore, after dividing the interval $[0,t]$ into $N$ equal
sub-intervals, $[j t/N,(j+1) t/N]$ one gets
\begin{equation}
  \label{eq:3060}
  V^tg(x,v) - \tV_{\epsilon}^tg(x,v) =
  \sum_{j=0}^{N-1} \tV_{\epsilon}^{j t/N} \left(
    V^{t/N}-\tV_{\epsilon}^{t/N}\right)V^{(N-1-j)t/N}g(x,v)\,. 
\end{equation}
The two semigroups are contractions  in $L^1\cap L^{\infty}$, and
therefore it is enough to prove that for $N=N_{\epsilon}$ 
appropriately chosen
\begin{equation}
  \label{eq:3065}
  N_{\epsilon} \left\|
    V^{t/N_{\epsilon}}g(x,v)-\tV_{\epsilon}^{t/N_{\epsilon}}g
  \right\|_{L^1}\rightarrow   0 
\end{equation}
when $\epsilon\rightarrow 0$. Setting $N_{\epsilon}=t/\taueps$ for a
suitable $\taueps$,  we find
\begin{align}
  & \left| V^{\taueps}g(x,v) - \tV_{\epsilon}^{\taueps}g(x,v) \right| =
  \left| \sum_{k=0}^{\infty}\left({(V^{\taueps})}_{k}
    g(x,v) - {(\tV_{\epsilon}^{\taueps})}_{k}  g(x,v)\right)
    \right|  \nonumber \\
  &\qquad\qquad\le
  \left| (V^{\taueps})_0 g(x,v) -
   (\tV_{\epsilon}^{\taueps})_0 g(x,v) \right| +
    \left| (V^{\taueps})_1g(x,v) -
    {(\tV_{\epsilon}^{\taueps})}_1g(x,v) \right| +
    \nonumber \\
  \label{eq:2070x}
   &\qquad\qquad\qquad + \left| \sum_{k=2}^{\infty} (V^{\taueps})_{k}
    g(x,v) \right|
    +  \left| \sum_{k=2}^{\infty} {(\tV_{\epsilon}^{\taueps})}_{k} g(x,v)\right|\,.   
\end{align}
Denote the four terms in the right-hand side by  $R_{I}, R_{II},
R_{III}$, and $R_{IV}$, and set
$r_{\epsilon}=\epsilon^{(2\nu-1)/2}\left(1+\log(t/\sqrt{\epsilon})\right)$.
From  Proposition~\ref{prop:pathlength} it
follows  that
the first term, accounting for paths with no jump in the given
interval, satisfies
\begin{equation}
  \label{eq:thm3_5}
  N_{\epsilon}\int_{\R^2\times S^1} R_I(x,v)\,  dxdv \le
  \frac{t}{\tau_{\epsilon}}
    C M^2  r_{\epsilon}^{1/2}\,,
  \end{equation}
  which converges to zero with $\epsilon$ if
  $r_{\epsilon}^{1/2}/\tau_{\epsilon}$ does. The second term accounts
  for  paths with 
exacly one jump, and according to Proposition~\ref{prop:propV1},
\begin{equation}
  \label{eq:thm3_6}
  N_{\epsilon}\int_{\R^2\times S^1} R_{II}(x,v)\,  dxdv \le
  C M^2 \left( \omega(\tau_{\epsilon},g) + r_{\epsilon} +
    \|g\|_{\infty} \sqrt{r_{\epsilon}}
    + \|g\|_{\infty} r_{\epsilon}
    /\tau_{\epsilon}\right) \,.
\end{equation}
For this term it is enough that $r_{\epsilon}/\tau_{\epsilon}$
converges to zero, but the rate of convergence may depend on the
modulus of continuity of the test function $g$.

We also have that if $\tau_{\epsilon} > r_{\epsilon}^{1/2} $
\begin{equation}
  \label{eq:thm3_7}
  N_{\epsilon}\int_{\R^2\times S^1} \left( R_{III}(x,v)+R_{IV}(x,v)\right)  \,
  dxdv
  \le C M^2 \|g\|_{\infty}  t \tau_{\epsilon}\,.
\end{equation}
This follows because for the Boltzmann process, the probability of
having more than two jumps in an interval of length $\tau_{\epsilon}$
is of the order $\tau_{\epsilon}^2$, and Proposition~\ref{prop:V2}
states that the same is true for the Markovian Lorentz process.
Therefore, in conclusion, the convergence stated in
eq.~(\ref{eq:3050}) holds with a rate depending on $f_0$ but
which can be controlled by entropy and moments, and on the modulus of
continuity of the test function $g$.

 \hfill $\square$

\medskip
 
The expression~(\ref{eq:3060}) implies that it is enough to study the
processes in very short intervals, and Proposition~\ref{prop:V2} below
shows that it is then enough to consider two cases: a particle
path starting at $(x,v)\in \R^2\times S^1$, {\em i.e.} with position
$x\in\R^2$ in the direction $v$, moves without changing velocity
during the whole interval, or, hits an obstacle at some point $x'$ and
continues from there in the new direction $v'$, but suffers no more
collisions. The three propositions used in the proof of
Theorem~\ref{thm:MarkovToBoltzmann} all depend on
Lemma~\ref{lem:onepassage} below, were the underlaying periodic
structure is used to analyze the particle paths in detail up to and
just after the first 
collision with an obstacle. To set the notation for the following
results, we  consider a path starting in the direction $v$ from a
point $x$ in the direction of $v$. It is sometimes  convenient to
denote the velocity by $v\in S^1$ by an angle $\beta\in [0,2\pi[$, and
both notations are used below without further comment. When a
collision takes place, the the new velocity $v'$ depends on $v$ and on
the impact parameter $r$ as shown in Fig.~\ref{fig:fig4}, where
$r\in[-1,1]$ when scaled with respect to the obstacle radius. In this
scaling, $dr = \cos(\beta'/2)\,d\beta'/2$, when $\beta'$ is given as
the change of direction.  

\begin{figure}[h]
  \centering
  \includegraphics[height=0.4\textwidth]{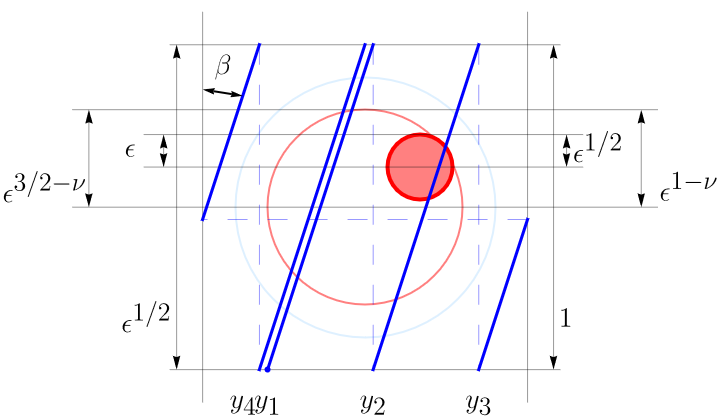}
  \caption{A particle path traversing the lattice cells. The cell
    size, diameter of the distribution of the obstacle center  and
    obstacle size are given 
    in the macroscopic scale to the left and microscopic scale to the right}
  \label{fig:fig3}
\end{figure}

\begin{figure}[h]
  \centering
  \includegraphics[height=0.4\textwidth]{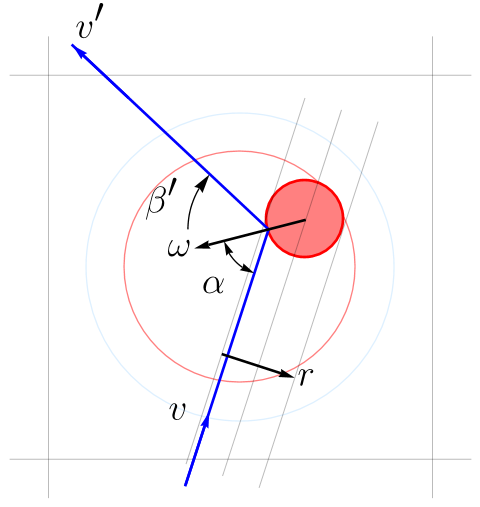}
  \caption{The figure illustrates a collision with an obstacle and the
  notation used to describe this.}
  \label{fig:fig4}
\end{figure}

Without loss of generality we may assume that the path is in the
upward direction 
with an angle $\beta\in [0,\pi/4] $ to the vertical axis as in
Fig.~\ref{fig:fig3}, because all other cases can be treated in the
same way just by finite number of rotations and reflections of the
physical domain. Let $y_1$ be the first time the path enters the lower
boundary of the lattice cell, and then let $y_j= y_1+(j-1)
\tan(\beta) \mod 1 $ be the consecutive points of entry to the lattice cell.
Setting $y_0=-\tan(\beta)/2$, the signed distance between the particle
path through the cell and the cell center is $\rho_j=(y_j-y_0)\cos\beta$,
and the probability that the path is reflected at the $j$-th passage
and given the sequence $\rho_j$, the events of scattering in cell
passage nr $j$ are independent.

Define
\begin{align}
  p_0(x,v,t) = \Pe\Big[ &\mbox{ no collision in the interval }
                          [0,t[ \nonumber\\
  &\qquad\qquad \mbox{ for a path starting at } (x,v)\in\R^2\times S^1
    \Big]\,.
\end{align}
The probability that a scattering event takes place in the $j$-passage
of a lattice cell is a function of the distance from the path to the
centre of the cell, $\rho_j$, and is zero outside
$|\rho_j|<\epsilon^{1-\nu} + \epsilon^{1/2}$.  This probability will be
denoted $p_j=p_j(x,v)$, and depends only on the initial position $x$
and the direction $v$.  Given that a
scattering event takes place, the outcome of this event depends on the
scattering parameter, {\em i.e.} the distance between the path and the
(random) center of the obstacle. Let $A_{x,v,t}$ be any event that
depends on a position $x$ and direction $v$ of a path segment of
length $t$. Then the probability that $A=A_{x',v',t'}$ is realized after
the first collision can
be computed as 
\begin{align}
  \label{eq:z00}
  \Pe[A] = \sum_{j=1}^n p_0(x,v,t_{j-})p_j(x,v) \Pe[ A_{x',v',t'} \; | \; j]\,.
\end{align}
Here $\Pe[ A_{x',v',t'} \; | \; j\;]$ denotes the conditional probability
of the event $A_{x',v',t'}$ given that the collision takes place when
the particle crosses cell number $j$ along the path, and this also
depends on the $x$ and $v$. The time when the path enters cell number
$j$ is denoted  $t_{j-}$, and the terms in the sum contains a factor
$p_0(x,v,t_{j-})$ for the probability that no collision has taken
place earlier. Similarly $t_{j+}$ denotes the time when the trajectory
leaves the cell; that is a random number, but we always have
$0< t_{j+}-t_{j-1} < 2\sqrt{\epsilon}$. 

In~(\ref{eq:z00}), $(x',v',t')$ are random, so
$\Pe[A]$ involves also an integral over these variables, and the
notation  $\Pe[ A_{x',v',t'} \; | \; j]$ is intended to include this
integration. The expectation $\EE[\psi]$ of some function $\psi$ is
computed in the same way with  $\Pe[ A_{x',v',t'} \; | \; j]$ replaced
by $\EE[ A_{x',v',t'} \; | \; j]$\,.

 Given the density $\phi$ for the position of the obstacle define
\begin{equation}
  \label{eq:z1}
  \varphi_0(x_1)=\int_{\R} \phi_0(x_1,x_2)\,dx_2\,,
\end{equation}
where $(x_1,x_2)$ is an arbitrary coordinate system in $\R^2$. Because
$\phi$ is assumed to be rotationally symmetric, the resulting function
$\varphi_0$ is independent of the orientation of this coordinate
system. It is a smooth probability density  with support in $[-1,1]$,
and then $\varphi_{\epsilon}(r) \equiv \epsilon^{\nu-1} \varphi_0(
\epsilon^{\nu-1}r)$ is also a smooth probability density with support
in $|r|\le\epsilon^{1-\nu}$.

For each $k$ the obstacle center is chosen
randomly inside a circle of radius $\epsilon^{1-\nu}$ around the cell
center, and one can then compute the probability that the path is
reflected at step $k$.

\begin{lm}
  \label{lem:onepassage}
  Consider a lattice cell in the microscopic scale, so that the cell
  side has length one, and assume that there is an obstacle of radius
  $\epsilon^{1/2}$ with center distributed with a density of the form 
  $\phi(x)=\epsilon^{2\nu-2} \phi_0( \epsilon^{\nu-1}x )$, where
  $\phi_0\in C^{\infty}(\R^2) $ has support in the unit ball of  $\R^2$.
   Let $\varphi_{\epsilon}$
  be defined as in Eq.~(\ref{eq:z1}) and rescaled with $\epsilon$.
  Let  $\psi_0\in L^1(\R)$ with $\psi_0(x)=0$ for $|x|>1$, and
  set $\psi_{\epsilon}(x) = \psi_0(x/\sqrt{\epsilon})$. 
  Consider a particle path traversing the cell $n$
  times with an angle $\beta\in [0,\pi/4]$ to the vertical cell sides,
  and set $y_1,...,y_n$ to be the consecutive points at the lower edge
  of the cell. We have $y_1\in [-1/2,1/2[$ and
  $y_k=-1/2 + ( 1/2 + (k-1)\tan(\beta)  \mod \; 1) $. 
 Then 
  \begin{equation}
    \label{eq:2075}
    p[\psi_{\epsilon}](y,\beta) = \int_{\R}
    \psi_{\epsilon}(x_1 )
    \phi_{\epsilon}(y\cos(\beta) +x_1)\,dx_1 
  \end{equation}
  satisfies
  \begin{equation}
    \label{eq:2076}
    \sum_{k=1}^n p[\psi_{\epsilon}](y_k,\beta) = \frac{\sqrt{\epsilon}
      \,n}{\cos(\beta)} \int_{1}^{1} \psi_0(x)\,dx + R_a(y_1,\beta)\,. 
  \end{equation}
  The remainder term $R_a$ also depends on $n$ and $\epsilon$.  It is
  bounded by a    function $\bar{R}_A(\beta)$, that depends on
  $\phi$, $n$ and  
  $\epsilon$ but not  on $\psi$, and that satisfies  
  \begin{align}
    \label{eq:2077}
    \left|R_a(n,\epsilon, y_1,\beta)\right| & \le \| \psi_0\|_{L^1}
                                              \bar{R}_A(\beta) 
                                              \qquad \mbox{where}\\
    \label{eq:2077b}
    \int \bar{R}_A(\beta) \, d\beta & \le
   C \epsilon^{\nu-1/2} (1+\log n)\,.
  \end{align}

\end{lm}

\noindent{\bf Proof:} The result is a small extension of the
corresponding proposition in~\cite{RicciWennberg2004}, and also the
proof follows closely that paper. For any fixed $\beta$, the expression  
$p[\psi_{\epsilon}](y,\beta)$ has support in an interval of length $2(
  \epsilon^{1-\nu} + \epsilon^{1/2})/\cos(\beta)$. Extend this
  function to be a one-periodic function of $y$. For simplicity we
  assume that $n$ is odd, and set $n=2m+1$. With $n$ large, this
  assumption would only make a  very small contribution from adding
  one extra term to the sum in case
  $n$ were even. Then 
  \begin{align}
    \label{eq:z10}
    \sum_{j=1}^n p[\psi_{\epsilon}](y_j,\beta) 
    &= \sum_{j=-m}^m p[\psi_{\epsilon}](y_1+
      (j+m-1)\tan(\beta),\beta)\nonumber\\
    &= \sum_{k=-\infty}^{\infty} \hat{p}_k \sum_{j=-m}^m e^{2\pi i k
      (y_1+(j+m-1)\tan(\beta))}\nonumber \\
    &=  \sum_{k=-\infty}^{\infty} \hat{p}_k e^{2\pi i k
      (y_1+(m-1)\tan(\beta))} \sum_{j=-m}^m e^{2\pi i k
      j\tan(\beta)}\,.
  \end{align}
In this expression
$\hat{p}_k$ is the
$k$-th Fourier coefficient of the periodic function
$p[\psi_{\epsilon}](\cdot,\beta)$, and the sum to the right can be
evaluated as the Dirichlet kernel of order $m$ with argument
$k\tan(\beta)$:
\begin{equation}
  \label{eq:z12}
  D_m(x) = \frac{\sin((2m+1) \pi x)}{\sin(\pi x)}\,.
\end{equation}
Therefore
\begin{equation}
  \label{eq:z14}
  \sum_{j=1}^n p[\psi_{\epsilon}](y_j,\beta) = (2m+1) \hat{p}_0 + R_a\,
\end{equation}
where
\begin{equation}
  \label{eq:z16}
  R_a = \sum_{k\ne 0} \hat{p}_k e^{2\pi i k
      (y_1+(m-1)\tan(\beta))} D_m(k\tan{\beta})\,.
\end{equation}
The Fourier coefficients $\hat{p}_k$ can be computed as the Fourier
transform of $p[\psi_{\epsilon}](\cdot,\beta)$ evaluated at the
integer points $k$,
\begin{equation}
  \label{eq:z16x}
  \hat{p}_k = \int_{-\infty}^{\infty} e^{-2\pi i k y}
  p[\psi_{\epsilon}](y,\beta)\,dy =
  \frac{\sqrt{\epsilon}}{\cos(\beta)}\widehat{\psi}_0(-\sqrt{\epsilon}k)
\widehat{\varphi}_0(\epsilon^{1-\nu}k)\,.
\end{equation}
Here $\widehat{\varphi}_0$ and $\widehat{\psi}_0$ are the Fourier transforms of
$\varphi_0$ and $\psi_0$. Note the factor $\sqrt{\epsilon}$ which is due
to the definition of $\psi_{\epsilon}$, which is not rescaled to
preserve the $L^1$-norm. We have
\begin{equation}
  \label{eq:18}
  \hat{p}_0 = \frac{\sqrt{\epsilon}}{\cos(\beta)}\int \psi_0(x)dx\,, 
\end{equation}
and because $\psi_0\in L^1$ and $\varphi_0$ is smooth, the coefficients
$\hat{p}_k$ decay rapidly: For any $a>0$ there is a constant $c_a$
depending on $\varphi_0$ such that
\begin{equation}
  \label{eq:20}
  |\hat{p}_k| \le  \sqrt{\epsilon} \| \psi_0\|_{L^1} \frac{c_a}{1+
    \left|\epsilon^{1-\nu}k\right|^a}\,. 
\end{equation}
Because $\beta\in[0,\pi]$, the dependence on $\beta$ can be absorbed
into the constant $c_a$.
Let
\begin{equation}
  \label{eq:barRdef}
  \bar{R}_A(\beta) = \sum_{k\ne 0}
  \frac{c_a\sqrt{\epsilon}}{1+\left|\epsilon^{1-\nu}k\right|^1}
  \left|D_m(k\tan\beta)\right| 
\end{equation}
The remainder term $R_a=R_a(n,\epsilon, y_1,\beta)$ is
therefore bounded by
\begin{equation}
  \left| R_a\right| \le \|\psi_0\|_{L^1} \bar{R}_A(\beta)\,.
\end{equation}
Using a standard estimate of the $L^1$-norm of the Dirichlet kernel, 
\begin{align}
  \label{eq:4170}
  \int_{0}^{\pi/4} | D_m(k \tan\beta) |\,d\beta
  & = \int_{0}^{1}
    \frac{\left|D_m( k \xi)\right|}{1+\xi^2}\,d\xi \nonumber \\
  & \le
  \frac{1}{k}\int_{0}^{k}\left|D_m(\xi)\right| \,d\xi\,\le\, C(1+\log m).
\end{align}
 Therefore these integrals are independent of $k$,
and so
\begin{align}
  \label{eq:20x1}
  \int_{S_1} \left| \bar{R}_A(\beta)\right|\,d\beta
  &\le C \sqrt{\epsilon} (1+\log(m)) \sum_{k\ne 0}\frac{c_a
    \|\psi_0\|_{L^1}}{1+ \left|\epsilon^{1-\nu}k\right|^a} \nonumber \\
  &\le C \|\psi_0\|_{L^1} \epsilon^{\nu-1/2} (1+\log(n) )\,,
\end{align}
because the sum is bounded by $C\int_{0}^{\infty} (1+
(\epsilon^{1-\nu}x)^a)^{-1}\,dx$, and replacing $\log(m)$ by $\log(n)$
only modifies the constant. This concludes the proof.

\hfill$\square$.

\begin{remark}
  The assumption that $\psi_0\in L^1$ is actually much stronger than
  needed. The proof works equally well as long as the Fourier
  transform of $\psi_0$ is not increasing too fast, and a
  Dirac mass with support in $]-1,1[$, for example,  would give the same kind of
  error estimate. 
\end{remark}

\begin{prop}
  \label{prop:pathlength}

  The terms $(V^t)_0 g (x,v)$ and $(\tV^t_{\epsilon})_0 g(x,v)$ are
  given by 
  \begin{align}    
  \label{eq:prop:pathlength1}
  {(V^t)}_0 g (x,v) & = e^{-2t} g(x+vt,v)\qquad \mbox{and}\\
  \label{eq:prop:pathlength2}
  {(\tV^t_{\epsilon})}_0 g(x,v) & = p_0(x,v,t)  g(x+vt,v)\,.
  \end{align}
   where $p_0(x,v,t)$ is the probability that a trajectory starting at
   $x\in\R^2$ in direction  $v$ does not hit an obstacle in the
   interval $[0,t]$; this can be computed explicitly. The function
   $p_0(x,v,t)$ satisfies $p_0(x,v,t)-e^{-2t} = R_b(x,v,t)$ with 
   \begin{align}
     \label{eq:p0estimate}
     |R_b(x,v,t)| & \le \bar{R}_B(\beta)
   \end{align}
   for a function $\bar{R}_B(\beta)$ that satisfies
   \begin{align}
     \int_{S_1} \bar{R}_B(\beta)  \, dv
     &\le C \epsilon^{2\nu-1)/4}\sqrt{1+\log(t/\sqrt{\epsilon})}\,,
    \end{align}
  and consequently, for any  $g\in
  C\left(\R^2\times S^1\right)$ with support in $|x|\le M$,  
  \begin{equation}
     \label{eq:3080}
     \left\|{(V^t)}_0 g - {(\tV^t_{\epsilon})}_0
       g\right\|_{L^1(\R^2\times S^1)}\le
     C M^2 \|g\|_{L^{\infty}}
     \epsilon^{(2\nu-1)/4}\sqrt{1+\log(t/\sqrt{\epsilon})}\,.
   \end{equation}
\end{prop}


\noindent{\bf Proof:} Consider a path starting at $x\in\R^2$ in the
direction of $v$. There is no restriction in assuming that the direction
$v$ is clockwise rotated with an angle $\beta$ as illustrated in
Fig.~\ref{fig:fig3}, which shows one lattice cell, with an obstacle
patch indicated with a red circle, and a blue circle indicating the
maximal range for an obstacle, and the red solid disk shows one
possible position of the obstacle. In the macroscopic scale, the
lattice size is $\epsilon^{1/2}$, the obstacle has radius $\epsilon$,
and the obstacle patch has radius $\epsilon^{3/2-\nu}$. These values
are indicated within parenthesis, and the microsopic scale, in which
the lattice size is $1$ is indicated to the left. When
$\epsilon\rightarrow0$ the obstacle patch in microscopic scale shrinks
to a point, but is is drawn large in the image for clarity. The path
under consideration enters a new lattice cell for the first time at $y_1$, and
enters the second lattice cell at $y_2$, drawn in the same image. A
path of length $t$ in macroscopic scale will traverse a number $n$ of lattice
cells in the vertical direction. We set $m =\lfloor \frac{t
  \cos\beta}{2\epsilon^{1/2}}\rfloor$. Then 
\begin{equation}
  \label{eq:3085}
  n = 2m+1 = \frac{t \cos \beta}{\epsilon^{1/2}} + \zeta\,,
\end{equation}
where $\zeta\in]-1,1]$, and depending on whether where the path starts
 and ends, the path may touch one additional cell. The error coming
 from not knowing the start and end points can be taken into account
 by allowing $\zeta\in[-2,2]$. Then
\begin{align}
  \label{eq:3090}
  p_0(x,v,t)&=\Pe[\;\mbox{no collision in the interval }\; [0,t[\;]
        \nonumber \\
  &= \prod_{j=1}^n (1-p(y_j,\beta)) = \exp\bigg(\sum_{j=1}^n
    \log(1-p(y_j,\beta))\bigg) 
\end{align}
where $p(y,\beta)$ is the probability that a trajectory entering a lattice
cell at $y$ with angle $\beta$ along the lower cell boundary meets an
obstacle before leaving the cell on the top. Because $p(y_j,\beta)<
c\epsilon^{\nu-1/2}\rightarrow 0$ when $\epsilon\rightarrow 0$ we
may assume that $p(y_j,\beta)<1/2$, and therefore
\begin{equation}
  \label{eq:4100}
  -(1+c \epsilon^{\nu-1/2}) p(y_j,\beta)< \log(1-p(y_j,\beta)) <
  -p(y_j,\beta)\,, 
\end{equation}
which provides an asymptotic expression for $p_0(x,v,t)$ once the sum
$\sum_{j=1}^n p(y_j,\beta)$ has been evaluated.
That $(V^t)_0 g(x,v)$ has the form (\ref{eq:prop:pathlength1}) is
evident, and hence it remains to prove the
estimate~(\ref{eq:3080}). From~(\ref{eq:4100}) we get
\begin{align}
  \label{eq:4180}
  \left|e^{-2t}-p_0(x,v,t)\right|
   &\le \left|e^{-2t} -e^{\sum_{j=1}^n p(y_1\beta) }\right| +
  \left|e^{-(1+c\epsilon^{\nu-1/2})\sum_{j=1}^n
     p(y_1,\beta)}-e^{-\sum_{j=1}^n p(y_j,\beta)}\right|
  \nonumber \\
  &\le \left|e^{-2t} -e^{-\sum_{j=1}^n p(y_1,\beta) }\right| +
    c\epsilon^{\nu-1/2}\,. 
\end{align}
Referring to the notation in Lemma~\ref{lem:onepassage} the
probability $p(y_j,\beta)$ can be computed as
\begin{equation}
  p(y_j,\beta) = p[\one_{\epsilon}](y,\beta)\,,
\end{equation}
that is, $\psi_0$ is set to one in the interval $[-1,1]$ and zero
outside this interval. The same lemma then gives
\begin{equation}
  \label{eq:pp2}
  \sum_{j=1}^n p(y_1,\beta) = \frac{2\sqrt{\epsilon}n}{\cos(\beta)} +
  R_a(n,\epsilon,y_1,\beta) = 2 t + \hat{R}_a\,,
\end{equation}
where $R_a$ satisfies the estimate~(\ref{eq:2077b}), and
$|\widehat{R}_a-R_a|\le 2 \sqrt{\epsilon}$. 
Hence  $|\hat{R}_a(n,\epsilon,y_1,\beta)| \le C
\bar{R}_A(\beta)$ for some constant $C$, and Markov's inequality says
that  for any $\lambda>0$,  the inequality  $m( \{v\in S^1 \,|\,
\bar{R}_A>\lambda\}) \le \frac{C}{\lambda}
\epsilon^{\nu-1/2}(1+\log(t/\sqrt{\epsilon}) $ holds. 
Therefore
\begin{align}
  \label{eq:4190}
  \int_{S_1}& \left\|e^{-2t}-p_0(\cdot,v,t)\right\|_{L^{\infty}}\,dv
  \le \int_{S^1}  \left\|e^{-2t} -e^{-2t + \hat{R}_3
              }\right\|_{L^{\infty}}\,dv + 2\pi c 
    \epsilon^{\nu-1/2}\nonumber \\
  &\le C \left( m( \{v\in S^1 \,|\, \bar{R}_A >\lambda\}) + \lambda +
    \epsilon^{\nu-1/2}\right) \le \nonumber \\
  &\le C    \frac{\epsilon^{\nu-1/2}(1+\log(t/\sqrt{\epsilon}))}
    {\lambda}+C \lambda + C \epsilon^{\nu-1/2} \le C 
    \epsilon^{(2\nu-1)/4} \sqrt{ 1+\log(t/\sqrt{\epsilon})}\,.
\end{align}
where the norms inside the integrals are taken with respect to the
variable $x$. Then (\ref{eq:3080}) follows after integrating the 
expression over 
$\R^2$ because
\begin{equation}
  \label{eq:4191}
  \left\|{(V^t)}_0 g - {(\tV^t_{\epsilon})}_0
    g\right\|_{L^{1}(\R^2\times S^1)} \le
  \|g\|_{L^{\infty}(\R^2\times S^1)} \pi M^2 \int_{S^1}\left\|  p_0(\cdot,v,t) -
    e^{-2t}\right\|_{L^{\infty}}\,dv\,, 
\end{equation}
when $g=0$ for $|x|>M$.

\hfill$\square$

\begin{prop}
  \label{prop:V2}

Let $t>\epsilon^a$ with $a<(2\nu-1)/4$. Then if $g=0$ for $|x|>M$,
  
\begin{equation}
    \left\| \sum_{k=2}^{\infty} {(\tV_{\epsilon}^{t})}_{k}g
    \right\|_{L^1(\R^2\times S^1)}\, \le  C \|g\|_{L^{\infty}} M^2 t^2\,.
  \end{equation}
  
\end{prop}

\noindent{\bf Proof:}
Let $B_M$ be the ball of radius $M$ in $\R^2$, and
consider a path starting at $(x,v)\in B_M \times S^1$, and set
$J_k=J_k(x,v,t)$ be the event that this path has
exactly $k$ velocity 
jumps in the time interval $[0,t]$. Similarly, let $J_{k+}$
denote the same for the case of  at least $k$ jumps. Then 
\begin{equation}
  \label{eq:b65}
   \left\| \sum_{k=2}^{\infty} {(\tV_{\epsilon}^{t})}_{k}g
   \right\|_{L^{1}(\R^2\times S^1)}
   \le \|g\|_{L^{\infty}(\R^2\times S^1)}
   \int_{B_M \times S^1} \Pe[J_{2+} \; | \; (x,v)]\,dx dv\,,
 \end{equation}
  that is, the
 conditional probability that a path has at least two velocity jumps
 given the  initial position and velocity $(x,v)$ is integrated over
 $x$ and $v$. 
 Considering  $J_{2+}(x,v,t)$ for one octant of $S^1$ at a time, we
 may represent 
 $v$ by an angle $\beta\in[0,\pi/4]$ as in the proof of
 Lemma~\ref{lem:onepassage}. A path in the direction of $\beta$
 starting at $x$ will traverse (almost exactly) $n =
 t \cos(\beta)/\sqrt{\epsilon}$ if it is not reflected on an obstacle
   along the way, and so
   \begin{align}
     \label{eq:66.00}
  \Pe[J_{2+} \; | \; (x,v)]
     & = \sum_{j=1}^{n} p_0(c,v,t_{j-})  p_j
       \Pe[ J_{1+}(x',v',t-t_{j+})\; | \; (j,x',v') ]
\end{align}
that is, as a sum of terms conditioned on the event that the first
jump takes place at the $j$-th passage of a lattice cell.
Clearly
   \begin{align}
     \label{eq:66.001}
  \Pe[J_{2+} \; | \; (x,v)]
     & \le \sum_{j=1}^{n}   p_j
       \Pe[ J_{1+}(x',v',t)\; | \; (j,x',v') ]\,,
\end{align}
and the terms $ p_j \Pe[ J_{1+}(x',v',t)\; | \; (j,x',v') ]  $ can be
expressed as 
 $p[\psi_{\epsilon}](y,\beta)$ with
\begin{equation}
  \label{eq:p65}
  \psi_{\epsilon} =
  \Big(1-p_0(x',v'(r/\epsilon))\Big)\one_{|r|\le\epsilon}
\end{equation}
in Lemma~\ref{lem:onepassage}; $p_0(x',v'(r/\epsilon),t)$ is the
probability that there is no collision in a path of length $t$
starting at $x'$ in the direction of $v'$, and this direction is given
as the  outcome of a collision with an obstacle with impact parameter
$r/\epsilon$.  Using  Proposition~\ref{prop:pathlength} we find that
$\left | p_0(x',v'(r/\epsilon),t)- e^{-2t}\right| \le
\bar{R}_B(\beta'(r/\epsilon))$, where $\beta'$ is the angular direction
corresponding to $v'$.
 Rescaling $\psi_{\epsilon}$ gives
\begin{align}
  \label{eq:p65.0}
  \left| \int_{-1}^1 \psi_{0} \,dr - 2 \left(1-e^{-2t}\right) \right| 
  &\le   \int_{-1}^1
  \bar{R}_B( v'(r),t)\,dr \nonumber \\
  &= 
    \int_{S^1}  \bar{R}_B(\epsilon,v') \frac{\cos(\beta'/2)}{2}\,dv'\,,
\end{align}
where $\beta'$ is the scattering angle of $v'$ with respect to the
velocity before scattering, $v$. It follows that
\begin{equation}
\label{eq:p65.01}
\|\psi_{0} \|_{L^1} \le 2 (1-e^{-2t}) + C \epsilon^{(2\nu-1)/4}
\sqrt{1+\log(t/\sqrt{\epsilon})}\,.    
\end{equation}
Therefore, using Lemma~\ref{lem:onepassage}, the righthand side of
Eq.~\ref{eq:66.00} is bounded by
\begin{equation}
  \label{eq:p65.01x}
  \frac{\sqrt{\epsilon} n }{\cos(\beta)}\|\psi_0\|_{L^1} + C
  \|\psi_0\|_{L^1}\bar{R}_A(\beta)\,, 
\end{equation}
and integrating with respect to $v$ over $S^1$ and then $x$ over $B_M$, gives
\begin{align}
  \label{eq:c65}
  \|g\|_{L^{\infty}} M^2  C
  & \left( 1-e^{-2t} + r_{\epsilon}^{1/2}\right) \left(\frac{\sqrt{\epsilon
    }n}{\cos(\beta)}+ r_{\epsilon} \right) \nonumber \\
  &\le \|g\|_{L^{\infty}} M^2C \left( t^2 +t
    (r_{\epsilon}^{1/2}+r(t,\epsilon)) + r_{\epsilon}^{3/2}\right)
\end{align}
as a bound for the right hand side of Eq.~(\ref{eq:b65}). Here
$r_{\epsilon}= \epsilon^{(2\nu-1)/2} (1+\log(t/\sqrt{\epsilon})$, and
$t$ is assumed  to be small. The proof is concluded by comparing the terms when $t>\epsilon^{a}$
with $a< (2\nu-1)/4$.

\hfill$\square$

The following proposition concerns the terms
${\left(V^t\right)}_1g(x,v)$ and ${\left(\tV_{\epsilon}^t\right)}_1g(x,v)$
defined by 
\begin{equation}
  \label{eq:5020}
  {\left(V^t\right)}_1g(x,v) = e^{-2t}\int_{0}^{t} \int_{S^1_-} g(x+\tau v + (t-\tau)v',v')|v\cdot\omega|\,d\omega\,d\tau
\end{equation}
and 
\begin{equation}
  \label{eq:5030}
  {\left(\tV^t_{\epsilon}\right)}_1 g(x,v) =
  \EE\left[g(x+\ttau v+(t-\ttau) v',v') \one_{J_1(x,v,t)}
  \right]\,, 
\end{equation}
where as before, $J_1(x,v,t))$ is the event that there is exactly one
jump on the trajectory starting at $x$ in the direction of $v$. The
$\ttau\in]0,t[$ and $v'\in S^1$ are the  random jump time and velocity
of the particle after the jump. 

\begin{prop}
  \label{prop:propV1}
Let $\omega(\delta,g)$ be the modulus of continuity for $g$, {\em i.e.} a
function such that $|g(x,v)-g(x_1,v_1)|\le
\omega(|x-x_1|+|v-v_1|,g)$. Then
\begin{align}
  \label{eq:propV1}
  &\left\|{\left(V^t\right)}_1g- {\left(\tV^t_{\epsilon}\right)}_1 g
  \right\|_{L^1(\R^2\times S^1)} \nonumber \\
  &\qquad\qquad\le
C M^2 \left( t\omega(t,g) +r_{\epsilon}\omega(t,g) +
    \|g\|_{L^{\infty} } r_{\epsilon} +  \|g\|_{L^{\infty} }
    t r_{\epsilon}^{1/2}  \right)
\end{align}
where the support of $g$ is contained in $\{ (x,v)\,|\,|x|<M\} $
and $r_{\epsilon} =\epsilon^{\nu-1/2}(1+\log n)$.

\end{prop}

\noindent{\bf Proof:}
First, because $g$ is assumed to have compact support, the modulus of
continuity exists, and $\omega(\delta,g)\rightarrow 0$ when
$\delta\rightarrow 0$. Define $g_x(v)=g(x,v)$, {\em i.e.} a function
depending only on $v$ but with $x$ as a parameter. Because
$|v|=|v'|=1$, we then  have  $|g_x(v')-g(x+\tau v + (t-\tau)v',v')|
\le \omega(t,g)$, and therefore
\begin{equation}
  \label{eq:propVI_p1}
  \left| {\left(V^t\right)}_1g_x-
    {\left(V^t\right)}_1g\right| \le t
  \omega(t,g)\,, 
\end{equation}
where the factor $t$ multiplying $\omega(t,g)$ comes from the integral
in the definition of $V^t_1$. Similarly
\begin{equation}
  \label{eq:propVI_p2}
  \left| {\left(\tV^t_{\epsilon}\right)}_1g_x-
    {\left(\tV^t_{\epsilon}\right)}_1g\right| \le (1- p_0(x,v,t))
  \omega(t,g) \le \left(1-e^{-2t} + \bar{R}_A(\beta)  \right)\omega(t,g)\,,
\end{equation}
where $p_0$ and $\bar{R}_A$ are defined as above. It remains to
compare  
\begin{equation}
  \label{eq:propVI_3}
  {\left(V^t\right)}_1g_x(v) = t e^{-2t} \int_{S^1_-} g_x(v') |v\cdot
  \omega|\,d\omega
\end{equation}
and
\begin{equation}
  \label{eq:propVI_3x}
  {\left(\tV^t_{\epsilon}\right)}_1g_x(v) = \EE\left[g_x(v')
    \one_{J_1(]x,v,t)}\right].
\end{equation}
The operators ${\left(V^t\right)}_1$ and
${\left(\tV^t_{\epsilon}\right)}_1$ act only  on the variable $v$ of
$g_x$, $x$ being considered as a parameter, and it is possible to use
Lemma~\ref{lem:onepassage}. By conditioning on the event that
a velocity jump  takes place in the $j$-the passage of a cell, 
\begin{align}
  \label{eq:propVI_4}
   \EE\left[g_x(v')
  \one_{J_1(x,v,t)}\right] 
  & = \sum_{j=1}^n p_j \EE\left[ g(v') p_0(x,v,t_{j-})
    p_0(x_j,v',t-t_{j+})\big|  j\right ]  
\end{align}
where $t_{j-}$ and $t_{j+}$ denote the time points when the trajectory
enters and leaves cell number $j$ along the path, $x_j$ is the
(random) point of reflection of the trajectory inside cell $j$,
$v'$ the random new velocity, and finally, $\EE[ f | j ]$ is shorthand
for the expectation of $f$ subject to the event that the jump takes
place in cell number  $j$ along the path. Then
\begin{align}
  \label{eq:propVI_5}
   \EE\left[g_x(v')
  \one_{J_1(x,v,t)}\right] 
  = &e^{-2t} \sum_{j=1}^n p_j \EE\left[ g(v')\big|  j\, \right ] \nonumber  \\
   & +  \sum_{j=1}^n p_j \EE\left[ g(v') p_0(x,v,t_{j-1})
     \left( p_0(x_j,v',t-t_{j-})- e^{-2(t-t_{j+})} \right) \big|
     j\right ]  \nonumber \\ 
   & +  \sum_{j=1}^n p_j \EE\left[ g(v') \left( p_0(x,v,t_{j-1})-
     e^{-2t_{j-}}\right) e^{-2(t-t_{j+})}\big|  j\right ] \nonumber \\
   & +  e^{-2t} \sum_{j=1}^n p_j \EE\left[ g(v')
     \left(e^{2(t_{j+}-t_{j-})}-1)\right) \big|  j\right ]  
\end{align}

Using Lemma~\ref{lem:onepassage} with $\psi_0(r)=
g_x(v'(r))\one_{|r|\le1}$ one finds that  the first term  is
\begin{equation}
  \label{eq:propVI_6}
  t \int_{-1}^1 g_x(v'(r))\,dr + R_a(n,\epsilon,x,\beta)\,,
\end{equation}
with $R_a$ bounded by $\|g\|_{L^{\infty}}\bar{R}_A(\beta)$ and 
$\int_{S^1} \bar{R}_A(\beta) \,dv'\le C \epsilon^{\nu-1/2}
(1+\log(t/\sqrt{\epsilon}))$ .
The second term is bounded by
\begin{equation}
  \label{eq:propVI_7}
  \|g\|_{L^{\infty}} \sum_{j=1}^n p_j \EE\left[
    \bar{R}_B(\beta'(r/\epsilon))  \big|  j\right ] =
  \|g\|_{L^{\infty}}  t \| \bar{R}_B(\beta(\cdot) \|_{L^1} +
  \|g\|_{L^{\infty}}  R_b(n,\epsilon,x,v)\,,
\end{equation}
where, in the same way as before, $\left|R_b\right|\le \|
\bar{R}_B(\beta(\cdot) \|_{L^1} \bar{R}_A(\beta)$\,.
Similarly the third term is bounded by
\begin{align}
  \label{eq:propVI_8}
  \|g\|_{L^{\infty}} & \sum_{j=1}^n  \left| p_0(x,v,t_{j-1})-
     e^{-2t_{j-}}\right| p_j \le  C  \|g\|_{L^{\infty}}
                       \bar{R}_B(\beta)\left(t+ C 
                       \bar{R}_A(\beta)\right)\,, \nonumber \\
    &\le C\|g\|_{L^{\infty}} \,  t \bar{R}_B(\beta) + C
      \|g\|_{L^{\infty}} \, \bar{R}_A(\beta)\,,
\end{align}
and the last term by
\begin{equation}
  \label{eq:propVI_9}
  C \sqrt{\epsilon} \left(t+ \bar{R}_A(\beta)\right)\,.
\end{equation}
Adding the error estimates from (\ref{eq:propVI_p1}),
(\ref{eq:propVI_p2}), (\ref{eq:propVI_6}), (\ref{eq:propVI_7}),
(\ref{eq:propVI_8}),and  (\ref{eq:propVI_9}) and integrating over
$\beta$ gives

\begin{align}
   \label{eq:propVI_10}
 & \int_{S^1}\left|{\left(V^t\right)}_1g(x,v)-
   {\left(\tV^t_{\epsilon}\right)}_1 g(x,v) \right|\,dv \nonumber \\
  &\qquad \le C \left( t\omega(t,g) +r_{\epsilon} \omega(t,g) +
    \|g\|_{L^{\infty} } r_{\epsilon} +  \|g\|_{L^{\infty} }
    t \,r_{\epsilon}^{1/2}  \right)\,,
\end{align}
again writing $r_{\epsilon}$ for the expression $\epsilon^{\nu-1/2}
(1+\log n)$. This  concludes the proof, because the integration over the support of
$g$ in $x$ is trivial.

\hfill$\square$

\section{Equivalence of processes and propagation of chaos}
\label{sec:equivalence}

The Lorentz process and the Markovian Lorentz process are equivalent
on the set of trajectories that don't return to the same obstacle
patch, and the purpose of this section is to prove that the
probability that a path in the Markovian process returns to the same
patch vanishes in the limit as $\epsilon\rightarrow 0$. This is a
 geometrical construction which consists in estimating the
measure of the set $B_{\epsilon}\subset \R^2\times S^1$ of initial
values that result in a path returning to the same obstacle
patch. This subset depends on the realization of the obstacle
positions, but the estimates of the measure $m(B_{\epsilon})$ of this
set do not, and 
$m(B_{\epsilon})\rightarrow 0$ when $\epsilon\rightarrow0$. A very
similar calculation is then made to prove that the probability that a
pair of trajectories i the Markovian process meet in an obstacle range
also converges to zero when $\epsilon$ does. The same calculation
could be repeated for any number of simultaneous trajectories, and
from that one can conclude that propagation of chaos holds for this
system.

\begin{figure}[h]
  \centering
  \includegraphics[height=0.6\textwidth]{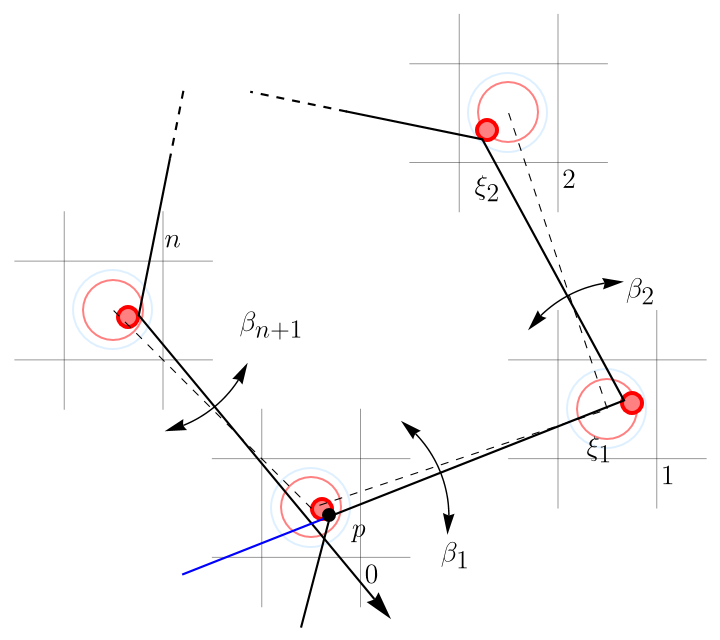}
  \caption{A path making a loop by returning to the same obstacle range
  in the lattice cell that here is indexed by $0\in\Z^2$}
  \label{fig:fig5}
\end{figure}

Consider then a path passing through an obstacle range, and then
returning to the same patch. All such paths can be enumerated by
giving the relative (integer) coordinates of the lattice cells where
the path changes direction, that is $\xi_j\in\Z^2$ denotes the
difference of the integer coordinates of the end point and starting
point of a straight line segment of the path. This is illustrated in
Fig.~\ref{fig:fig5}, where the starting point $p$ of the loop is
indicated with a black dot in cell nr. 0, which in this case is a
point where the path is reflected, but in most cases it would not be.
The point is not the starting point of the particle path, which may
have a long history before entering this particular loop. For a loop
with $n$ 
collisions, we therefore have
\begin{equation}
  \label{eq:5_00}
  \xi_1+\xi_2 + ... + \xi_{n+1} = 0 \in \Z^2\,.
\end{equation}
For a given time interval $t\in[0,\bart [$, we must have
\begin{equation}
  \label{eq:5_02}
  |\xi_i|+ |\xi_2| + \cdots + |\xi_{n+1}| \le \bart \epsilon^{-1/2}\,,
\end{equation}
because the size of a lattice cell is $\sqrt{\epsilon}$ in the scaling
we are considering here. The notation is indicated in
Fig.~\ref{fig:fig5}, where the first few and the last segment of a
loop are indicated with a solid line, the segments joining the
lattice centers with dashed lines, and angles $\beta_j$ of the path
segments, as well as the integer coordinates.

When $\epsilon$ is small, the path segments are almost parallel to
the corresponding segment joining the lattice cell centers, and the
same holds for the segment lengths. In the worst
case, with a segment joining two neighboring cells, the relative error
at most of order 
$\bigoh(\epsilon^{1-\nu})$. In the following computation, all lengths
are expressed in terms of the integer coordinates, and the error
coming from this approximation is taken into account by a constant
denoted $c$.

A loop returning to the same obstacle range is not uniquely determined
by the integer sequence $\xi_1,...,\xi_{n+1}$, because there is some
freedom in setting the initial  position and angle of the path. The last angle
$\beta_{n+1}$ must belong to an interval of width $\Delta \beta_{n+1}$
that satisfies
\begin{equation}
  \label{eq:5_10}
  \Delta \beta_{n+1} \le c \frac{\epsilon^{3/2-\nu}}
                         {\sqrt{\epsilon} |\xi_{n+1}|}\, 
\end{equation}
where the enumerator is the diameter of the obstacle range, and
$\sqrt{\epsilon} |\xi_{n+1}|$ is the length of the last segment. And it
is an easy argument to see that then the angle of the $n$-th segment
must belong to an interval of width smaller than
\begin{equation}
  \label{eq:5_11}
  \Delta \beta_{n} \le c \frac{\epsilon}{\sqrt{\epsilon} |\xi_n| }
  \Delta \beta_{n+1}\,.
\end{equation}
Following the path backwards gives
\begin{equation}
  \label{eq:5_12}
  \Delta \beta_1 \le c^n \frac{\epsilon^{n/2}}{|\xi_1| |\xi_2|\cdot
    \dots \cdot |\xi_n| }\Delta \beta_{n+1} \le
   c^{n+1} \frac{\epsilon^{n/2}}{|\xi_1| |\xi_2|\cdot
     \dots \cdot |\xi_n| } \frac{\epsilon^{1-\nu}}{|\xi_{n+1}|} \,.
\end{equation}
The history of the path leading up to the point $p$ is at most $\bart$,
and hence the phase space volume spanned by the possible histories
leading to a loop indexed by a sequence $(\xi_1,...,\xi_n)$ is bounded
by $ \bart c \epsilon^{3/2-\nu} \Delta \beta_1 $, the length of the path
times the diameter of of the obstacle range times the angular
interval. The constant $c$ is there to account for the difference
between the diameter of the obstacle range and obstacle patch, and can be
taken as close to 1 as one wishes. Of course the history is not likely
to be one straight line segment, but different histories leading a
particular loop may be very different, with none or many
reflections. However, it is well-known that the so called billiard map
is measure preserving. Let $\partial \Omega=\bigcup_{z \in
  \sqrt{\epsilon} \Z^2 } \{ x\in\R^2\;|\; x-z = \epsilon^{3/2-\nu}\}$, {\em i.e.}
    the union of all obstacle boundaries. If $(x,v)\in \partial \Omega
    \times S^1_+$, a point on the boundary of an obstacle, with
    velocity pointing out from the obstacle, then the billiard map is
    the map  $(x,v) \mapsto (x_1,v_1)$
where $(x_1,v_1)$ are the position and outgoing velocity after the
trajectory hits the next obstacle. This map preserves the measure
$  |n \cdot v| dx_{\parallel}dv$, where $n$ is the normal point out from
the obstacle at $x$ and $d x_{\parallel}$ is the length mesure of the
obstacle boundary, see e.g.~\cite{ChernovMarkarian2006}.  This means
exactly that even if the set of 
histories leading leading to the loop indexed by $(\xi_1,...,\xi_n)$
splits into a complicated form, the phase space volume in the full
space still satisfies the same bound. Because the loop is indexed with
coordinates relative to the starting point, all periodic translates of
a point of the history maps into a loop of with the same index
sequence, the fraction of points $(x,v)\in B\times S^1$, where $B$ is
any one lattice cell, that leads to
a loop with the given index is bounded by
\begin{align}
  \label{eq:5_18}
  \bart c \epsilon^{3/2-\nu} \Delta \beta_1 m(B)^{-1}
  &\le \bart c
  \epsilon^{3/2-\nu} c^{n+1} \frac{\epsilon^{n/2}}{|\xi_1| |\xi_2|\cdot
    \dots \cdot |\xi_n| } \frac{\epsilon^{1-\nu}}{|\xi_{n+1}|}
    \epsilon^{-1}\nonumber\\
  &=\bart c c^{n+1} \epsilon^{2(1-\nu) + (n-1)/2}
    \frac{1}{|\xi_1||\xi_1|\cdots|\xi_{n+1}|} \,.
\end{align}
To compute an estimate of the fraction of $(x,v)\in B\times S^1$ that
leads to any loop with $n+1$ segments it is enough to sum this
expression over the set of 
$(\xi_1,...,\xi_n,\xi_{n+1})$ satisfying
\begin{align}
  &\xi_1+\xi_2+ ... + \xi_{n+1}=0;\,,\qquad \xi_j\ne 0\,,\nonumber \\
  &  |\xi_1|+|\xi_2|+ ... + |\xi_{n+1}| \le \bart/\sqrt{\epsilon}\,. \nonumber 
\end{align}
Using $|\xi_1|^{-1}|\xi_{n+1}|^{-1}\le\frac{1}{2}
\left(|\xi_1|^{-2}+|\xi_{n+1}|^{-2}\right)$,  we find that the sum is
bounded by 
\begin{align}
  &\sum \frac{1}{|\xi_1|^2|\xi_2|\cdots|\xi_{n}|} \nonumber\\
  &\qquad \le C \int_{|x_1|\le \bart/\sqrt{\epsilon}}\frac{1}{|x_1|^2}
  dx_1\int_{|x_2|+...+|x_n|\le \bart/\sqrt{\epsilon}-|x_1|}
    \frac{1}{|x_2|\cdots|x_{n+1}|} dx_2\cdots dx_n \nonumber\\
  &\qquad\le C (2\pi)^n\int_1^{\bart/\sqrt{\epsilon}}
    \frac{1}{s}\,ds \;\frac{(\bart/\sqrt{\epsilon})^{n-1}}{(n-1)!}
    = C (2\pi)^n \log(\bart/\sqrt{\epsilon})
   \frac{\bart^{n-1}}{(n-1)!\epsilon^{(n-1)/2}}\,. 
\end{align}
The last line has been obtained by changing to polar coordinates in
$\R^2$, and taking the simplex in the inner integral to span over the
full length $T/\sqrt{\epsilon}$. Therefore, summing the estimate
in~(\ref{eq:5_18}) first over the loops of length $n+1$ and then over
$n=1\cdots\infty$ the following estimate for the fraction of initial
points in phase space that result in a loop along the path:
\begin{equation}
  \label{eq:5_25}
  \bart C e^{c \bart} \log(\bart/\sqrt{\epsilon}) \; \epsilon^{2(\nu-1)} \,.
\end{equation}
This calculation may be summarized as a proposition:
\begin{prop}
  \label{prop:noloops}
  Let $T_{\mathcal{Y}_{\epsilon,\epsilon}}^t$ be the Lorentz process as
    in Theorem~\ref{thm:main}. Denote the event that there is a loop
    along the path of length $\bart$ starting at $(x,v)$ by
    $\mathcal{L}(x,v)$ . Then 
    \begin{equation}
      \label{eq:5_30}
      \int_{\R^2\times S^1} f_0 (x,v)
      \EE\left[g(T^t_{\mathcal{Y}_{\epsilon,\epsilon}}(x,v))
          \one_{\mathcal{L}(x,v)}\right] \,dx dv \rightarrow 0
      \end{equation}
      when $\epsilon \rightarrow 0$
\end{prop}
\noindent{\bf Proof:}
It is enough to see that the integral is bounded by 
\begin{equation}
  \label{eq:5_26}
  \| g \|_{\infty}\int_{B_M\times S^1}f_0(x,v) \one_{\mathcal{L}(x,v)}\,dxdv\,  
\end{equation}
where $B_M$ is the ball of radius $M$ in $\R^2$, and assumed to
contain the support of $g$. This is bounded by
\begin{align}
  \label{eq:5_28}
  &\| g \|_{\infty}\int_{B_M \times S^1}f_0(x,v) \one_{\{f_0>\lambda\}}\,dxdv\, +
  \| g \|_{\infty}\lambda \int_{B_M \times
    S^1}\one_{\mathcal{L}(x,v)\cup \{ f<\lambda\} }\,dxdv\, \nonumber \\
  &\le \| g \|_{\infty}\int_{B_M \times S^1}f_0(x,v)
    \one_{\{f_0>\lambda\}}\,dxdv + \|g\|_{\infty} \lambda
    \bart C e^{c \bart} \log(\bart/\sqrt{\epsilon}) \; \epsilon^{2(\nu-1)}\,.
\end{align}
This expression can be made arbitrarily small by first choosing
$\lambda$ large enough to make the first term as one wishes, and then
the second term can be made equally small by choosing $\epsilon$
sufficiently small. All this is uniform in the random positions of the
obstacles. 

\hfill$\square$

The proof of propagation of chaos is similar in many ways. Consider
two paths of the Markovian Lorentz model with independent initial
conditions $(x,v)$ and $(x',v')$. Because there is no interaction
between the two particles, the particles remain independent until they
meet the same obstacle range, if ever. As in the previous calculation,
take a fixed realization of the random configuration of obstacles in
each obstacle range. Let $A(x,z)$ be the set of angles
$v\in S^1$ such that there is possible path from the point $x\in\R^2$
to the obstacle range with center at $z\in\sqrt{\epsilon}\Z^2$. The
set of angles such that the path reaches the patch before hitting an
obstacle is bounded by $\epsilon^{3/2-\nu} /|x-z|$, and the measure of
angles $v$ such that the path meets the obstacle patch after at least
one collision with an obstacle can be computed as above. The result is
that
\begin{equation}
\label{eq:5_30x}
  m(A(x,z)) \le C \epsilon^{3/2-\nu} \left( \frac{1}{|x-z|} +
    e^{C\bart}\log(\bart/\sqrt{\epsilon}) \right)\,,  
\end{equation}
and the set of angles $v$ and $v'$ such that both the path $T^t(x,v)$
and $T^t(x',v')$ meet the same obstacle range is bounded by
\begin{align}
  \label{eq:5_32}
  &m\left(\left\{ (v,v')\in (S^1)^2 \; |\; v\in A(x,z), v'\in A(x',z)
  \right\}\right) \nonumber \\
  &\qquad \le
  C \epsilon^{3-2\nu}\left( \frac{1}{|x-z|}+ e^{C\bart}
    \log(\bart/\sqrt{\epsilon})\right)
  \left( \frac{1}{|x'-z|}+ e^{C\bart} \log(\bart/\sqrt{\epsilon})\right)\,.
\end{align}
This expression should now be summed over all possible $z\in
\sqrt{\epsilon} \Z^2$, and because the speed of a particle is equal to
one, these $z$ belong to a ball of diameter $\bart$. As above we find
\begin{align}
  \label{eq:5_33}
  \sum_{z\in\Z^2, |z|<\bart/\sqrt{\epsilon}} \frac{1}{|x-z|}\frac{1}{|x-z|}
  &\le \frac{1}{2}\sum_{z\in\Z^2, |z|<\bart/\sqrt{\epsilon}}
  \left(\frac{1}{|x-z|^2}+\frac{1}{|x'-z|^2}\right) \nonumber\\
  &  \le C \log(\bart/\sqrt{\epsilon})\,,
\end{align}
and
\begin{align}
  \sum_{z\in\Z^2, |z|<\bart/\sqrt{\epsilon}} \left(\frac{1}{|x-z|}+\frac{1}{|x-z|}\right)
  &\le C \bart/\sqrt{\epsilon}\,.
\end{align}
The dominating term when summing the expression in~(\ref{eq:5_32})
comes from the part that  does not depend on $z$. We get
\begin{equation}
  \label{eq:5_35}
  \sum_{z\in\Z^2, |z|<\bart/\sqrt{\epsilon}} m\left(\left\{ (v,v')\in
      (S^1)^2 \; |\; v\in A(x,z), v'\in A(x',z) 
  \right\}\right) \le C(1+\bart^2)  e^{C \bart}
\log(\bart/\sqrt{\epsilon})  \epsilon^{2(1-\nu)}\,,   
\end{equation}
which again decreases to zero when $\epsilon$ does.
The maps $T_{\calye}^t$ and $\tT_{\calye}^t$ extend to pairs of
particles in a natural way, so that
\begin{align}
  (x(t),v(t),x'(t),v'(t))
  &= T_{\calye}^t( x_0,v_0,x'_0,v'_0)\;\mbox{and}\nonumber\\
  (\tilde{x}(t),\tilde{v}(t),\tilde{x}'(t),\tilde{v}'(t))
  &= \tT_{\calye}^t( x_0,v_0,x'_0,v'_0)
\end{align}
denote the position in phase space of a pair of particles evolving with
the Lorentz process and the Markovian Lorentz process respectively. In
the latter case the obstacle position inside an obstacle patch is
determined independently for the two particles, and every time a path
meets an obstacle range, and therefore
\begin{equation}
  \label{eq:5_40}
  \tT_{\calye}^t( x_0,v_0,x'_0,v'_0) = ( \tT_{\calye}^t( x_0,v_0),
  \tT_{\calye}^t(x'_0,v'_0))  \,.
\end{equation}
In the Lorenz evolution this is breaks down as soon as there is a
loop for one of the particle paths, or when the two particle paths
meet in one and the same obstacle range. However, the computation
above leads to the following theorem:

\begin{thm}
  \label{thm:chaos}
    Let $T_{\mathcal{Y}_{\epsilon,\epsilon}}^t$ be the Lorentz process as
    in Theorem~\ref{thm:main}. Then for $g_1, g_2\in C(B_M\times S^1)$
    \begin{align}
      \label{eq:5_50}
      \lim_{\epsilon\rightarrow 0}  
        \int_{(\R^2\times S^1)^2} f_0(x,v)& f_0(x',v') \Big(
        \EE\left[
          g_1\circ g_2 ( T_{\mathcal{Y}_{\epsilon,\epsilon}}^t(x,v,x',v'))
        \right] - \nonumber \\
        & \EE\left[
          g_1( T_{\mathcal{Y}_{\epsilon,\epsilon}}^t(x,v))
        \right]
        \EE\left[
          g_2( T_{\mathcal{Y}_{\epsilon,\epsilon}}^t(x',v'))
        \right]\Big)
        \,dx dv dx' dv'     = 0\,.
    \end{align}
  This says that the evolution of two particles in the Lorentz gas
  become independent in the limit as $\epsilon\rightarrow 0$.  
\end{thm}

\noindent{\bf Proof:}
Let $\mathcal{L}_2(x,v,x',v')$ denote the event that the two particle
paths starting at $(x,v)$ and $(x',v')$ and evolving in the same
random obstacle configuration meet in an obstacle range. Then
\begin{align}
  \label{eq:5_52}
  &\EE\left[
          g_1\circ g_2 (
          T_{\mathcal{Y}_{\epsilon,\epsilon}}^t(x,v,x',v'))
          (1-\one_{\mathcal{L}_2(x,v,x',v')})
  \right] \nonumber \\
  &\qquad\qquad =
    \EE\left[ g_1(T_{\mathcal{Y}_{\epsilon,\epsilon}}^t(x,v))
          g_2 (T_{\mathcal{Y}_{\epsilon,\epsilon}}^t(x',v'))
          (1-\one_{\mathcal{L}_2(x,v,x',v')})
        \right]
\end{align}
Therefore the integral in ~(\ref{eq:5_50}) is bounded by
\begin{equation}
  \label{eq:5_55}
  \|g_1\|_{\infty} \|g_2\|_{\infty}
  \EE\left[\int_{(\R^2\times S^1)^2} f(x,v) f(x',v')
    \one_{B(x,v,x',v')}\,,dxdvdx'dv'\right]\,. 
\end{equation}
Fix $\varepsilon>0$ arbitrary, and take $M>0$ and $\lambda>0$ so large
that
\begin{equation}
  \label{eq:5_56}
  \int_{\R^2\times S^1} f(x,v) (\one_{|x|>M}+ \one_{f>\lambda})\,dxdv
  < \varepsilon  \,.
\end{equation}
We find that~(\ref{eq:5_50}) is bounded by
\begin{equation}
  \label{eq:5_60}
  \|g_1\|_{\infty} \|g_2\|_{\infty} 2 \varepsilon +
  C(\bart) \lambda_2 R^2 \log(\bart/\sqrt{\epsilon}) \epsilon^{2(1-\nu)} 
\end{equation}
which can be made smaller then $2\varepsilon$ by choosing $\epsilon$
sufficiently small. Here $C(\bart)$ is a $\bart$ depending constant coming
from the expression~(\ref{eq:5_35}). This concludes the proof because
$\varepsilon$ was arbitrarily small.

\hfill$\square$

Theorem~\ref{thm:chaos} could have been proven for tensor products of
any order with exactly the same kind of computations, and therefore
this really is a proof that propagation of chaos holds. The rate of
convergence in the theorem depends on the density $f_0$, and also on
the time interval. The dependence of $f$ could have been replaced by
bounds of the moments and entropy. The dependence of the time interval
is much more difficult to get passed, but could maybe be addressed as
in~\cite{LutskoTot2020} which deals with the Lorentz gas in a Poisson
setting.

\bibliographystyle{plain}


\begin{thebibliography}{10}

\bibitem{BourgainGolseWennberg1998}
Jean Bourgain, Fran\c{c}ois Golse, and Bernt Wennberg.
\newblock On the distribution of free path lengths for the periodic {L}orentz
  gas.
\newblock {\em Comm. Math. Phys.}, 190(3):491--508, 1998.

\bibitem{CagliotiPulvirentiRicci2000}
E.~Caglioti, M.~Pulvirenti, and V.~Ricci.
\newblock Derivation of a linear {B}oltzmann equation for a lattice gas.
\newblock {\em Markov Process. Related Fields}, 6(3):265--285, 2000.

\bibitem{CagliotiGolse2010}
Emanuele Caglioti and Fran{\c{c}}ois Golse.
\newblock On the {B}oltzmann-{G}rad limit for the two dimensional periodic
  {L}orentz gas.
\newblock {\em J. Stat. Phys.}, 141(2):264--317, 2010.

\bibitem{ChernovMarkarian2006}
Nikolai Chernov and Roberto Markarian.
\newblock {\em Chaotic billiards}, volume 127 of {\em Mathematical Surveys and
  Monographs}.
\newblock American Mathematical Society, Providence, RI, 2006.

\bibitem{Gallavotti1972}
Giovanni Gallavotti.
\newblock Rigorous theory of the {B}oltzmann equation in the {L}orentz gas.
\newblock {\em Nota Interna, Istituto di Fisica, Universit\`{a} di Roma},
  (358), 1972.

\bibitem{Gallavotti1999book}
Giovanni Gallavotti.
\newblock {\em Statistical mechanics}.
\newblock Texts and Monographs in Physics. Springer-Verlag, Berlin, 1999.
\newblock A short treatise.

\bibitem{GikhmanSkorokhod1974}
Iosif~I. Gikhman and Anatoli~V. Skorokhod.
\newblock {\em The theory of stochastic processes. {I}}.
\newblock Classics in Mathematics. Springer-Verlag, Berlin, 2004.
\newblock Translated from the Russian by S. Kotz, Reprint of the 1974 edition.

\bibitem{Golse2008}
Fran\c{c}ois Golse.
\newblock On the periodic {L}orentz gas and the {L}orentz kinetic equation.
\newblock {\em Ann. Fac. Sci. Toulouse Math. (6)}, 17(4):735--749, 2008.

\bibitem{Lorentz1905}
Hendrik Lorentz.
\newblock Le mouvement des \'{e}lectrons dans les m\'{e}teaux.
\newblock {\em Arch. N\'{e}erl.}, 10:336--371, 1905.

\bibitem{LutskoTot2020}
Christopher Lutsko and B\'{a}lint T\'{o}th.
\newblock Invariance principle for the random {L}orentz gas---beyond the
  {B}oltzmann-{G}rad limit.
\newblock {\em Comm. Math. Phys.}, 379(2):589--632, 2020.

\bibitem{MarklofStrombergsson2008}
Jens Marklof and Andreas Str{\"o}mbergsson.
\newblock Kinetic transport in the two-dimensional periodic {L}orentz gas.
\newblock {\em Nonlinearity}, 21(7):1413--1422, 2008.

\bibitem{MarklofStrombergsson2010}
Jens Marklof and Andreas Str{\"o}mbergsson.
\newblock The distribution of free path lengths in the periodic {L}orentz gas
  and related lattice point problems.
\newblock {\em Ann. of Math. (2)}, 172(3):1949--2033, 2010.

\bibitem{MarklofStrombergsson2011b}
Jens Marklof and Andreas Str{\"o}mbergsson.
\newblock The {B}oltzmann-{G}rad limit of the periodic {L}orentz gas.
\newblock {\em Ann. of Math. (2)}, 174(1):225--298, 2011.

\bibitem{MarklofStrombergsson2011}
Jens Marklof and Andreas Str{\"o}mbergsson.
\newblock The periodic {L}orentz gas in the {B}oltzmann-{G}rad limit:
  asymptotic estimates.
\newblock {\em Geom. Funct. Anal.}, 21(3):560--647, 2011.

\bibitem{MarklofStrombergsson2014}
Jens Marklof and Andreas Str\"{o}mbergsson.
\newblock Free path lengths in quasicrystals.
\newblock {\em Comm. Math. Phys.}, 330(2):723--755, 2014.

\bibitem{MarklofStrombergsson2015}
Jens Marklof and Andreas Str\"{o}mbergsson.
\newblock Generalized linear {B}oltzmann equations for particle transport in
  polycrystals.
\newblock {\em Appl. Math. Res. Express. AMRX}, (2):274--295, 2015.

\bibitem{MarklofStrombergsson_arxiv2019}
Jens Marklof and Andreas Str\"{o}mbergsson.
\newblock Kinetic theory for the low-density {L}orentz gas.
\newblock arXiv:1910.04982, to appear in {\em Memoirs of the AMS}, 2019.

\bibitem{MelbournePeneTerhesiu2021}
Ian Melbourne, Fran{\c{c}}oise P{\`{e}}ne, and Dalia Terhesiu.
\newblock Local large deviations for periodic infinite horizon {L}orentz gases,
  2021.

\bibitem{PeneTerhesiu2020}
Fran{\c{c}}oise P{\`{e}}ne and Dalia Terhesiu.
\newblock Sharp error term in local limit theorems and mixing for {L}orentz
  gases with infinite horizon.
\newblock {\em Communications in Mathematical Physics}, 382(3):1625--1689, feb
  2021.

\bibitem{RicciWennberg2004}
Valeria Ricci and Bernt Wennberg.
\newblock On the derivation of a linear {B}oltzmann equation from a periodic
  lattice gas.
\newblock {\em Stochastic Process. Appl.}, 111(2):281--315, 2004.

\bibitem{Spohn1978}
Herbert Spohn.
\newblock The {L}orentz process converges to a random flight process.
\newblock {\em Comm. Math. Phys.}, 60(3):277--290, 1978.

\end{thebibliography}

\def\cprime{$'$} \def\cprime{$'$}
  \def\polhk#1{\setbox0=\hbox{#1}{\ooalign{\hidewidth
  \lower1.5ex\hbox{`}\hidewidth\crcr\unhbox0}}}

\end{document}